\documentclass[prd,showpacs,preprint,preprintnumbers,amsmath,amssymb,eqsecnum,nofootinbib]{revtex4-1}
\usepackage{graphicx,amssymb,amsmath,amsthm,amsfonts,mathrsfs,epsfig,epsf}
\usepackage{dcolumn}
\usepackage{amsfonts,mathrsfs}
\usepackage{bm}
\usepackage{subfigure}
\usepackage{color}
\usepackage{url}

\begin{document}

\preprint{RUP-18-34}

\title{Particle creation in gravitational collapse to a horizonless
compact object}

\author{Tomohiro Harada$^{1}$}
\email{harada@rikkyo.ac.jp}
\author{Vitor Cardoso$^{2,3}$}
\author{Daiki Miyata$^{1}$}
\affiliation{$^{1}$Department of Physics, Rikkyo University, Toshima,
Tokyo 171-8501, Japan}
\affiliation{$^{2}$CENTRA, Departamento de F\'isica, Instituto Superior T\'ecnico, Universidade de Lisboa,
Avenida Rovisco Pais 1, 1049 Lisboa, Portugal}
\affiliation{$^{3}$Theoretical Physics Department, CERN 1 Esplanade des Particules, Geneva 23, CH-1211, Switzerland}
\date{\today}
%
\begin{abstract}
Black holes (BHs) play a central role in physics. However, gathering observational evidence for their existence
is a notoriously difficult task. Current strategies to quantify the evidence for BHs all boil down to looking for signs of highly compact, horizonless bodies. Here, we study particle creation by objects which collapse to form
 ultracompact configurations, with the surface at an areal radius $R=R_{f}$
 satisfying $1-(2M/R_{f})= \epsilon^{2}\ll 1$ with $M$ the object
 mass. We assume that gravitational collapse proceeds in a ``standard''
 manner until $R=R_{f}+2M \epsilon^{2\beta}$, where $\beta>0$, 
and then slows down to form a static object of radius $R_{f}$. In the
 standard collapsing phase, Hawking-like thermal radiation is emitted, which is as strong as the Hawking radiation of a BH with the same mass but
lasts only for $\sim 40~(M/M_{\odot})[44+\ln (10^{-19}/\epsilon)]~\mu \mbox{s}$.
Thereafter, in a very large class of models, 
there exist two bursts of radiation separated by a very long dormant
 stage.
The first burst occurs at the end of the transient Hawking radiation
and is followed by a quiescent stage which lasts for  
$\sim 6\times 10^{6}~(\epsilon/10^{-19})^{-1}(M/M_{\odot})~\mbox{yr}$.
Afterwards, the second burst is triggered, after which there is no more 
particle production and the star is forever dark.  
In a model with $\beta=1$, 
both the first and second bursts outpower the transient Hawking radiation by a factor
 $\sim 10^{38}(\epsilon/10^{-19})^{-2}$. 
\end{abstract}


\maketitle

\tableofcontents

\section{Introduction and summary}
It is generally accepted that black holes (BHs) can be and have been found in various astrophysical systems, such as x-ray binaries,
galactic nuclei, and binary systems sourcing gravitational waves. These systems all contain dark, compact, and massive objects
whose properties are all consistent with the BH paradigm.
However, BHs are defined by the existence of an event horizon,
which is the boundary of the causal past of future null infinity. By its own definition, finding observational {\it proof} for event horizons is impossible~\cite{Abramowicz:2002vt,Cardoso:2017cqb,Cardoso:2016rao,Nakao:2018knn}. Thus, sufficiently compact bodies can mimic BHs at a classical level. Given the crucial role of horizons in a number of fundamental issues, quantifying the evidence for BHs is as important as quantifying, say, the level to which the equivalence principle is satisfied~\cite{Abramowicz:2002vt,Cardoso:2017cqb,Cardoso:2016rao,Cardoso:2017njb,Cardoso:2017cfl,Maselli:2017cmm,Shaikh:2018lcc,Cunha:2018gql}.

A natural strategy to test the BH paradigm is to look for smoking-gun imprints of {\it horizonless} bodies.
The number of proposals for ultracompact horizonless objects is large
and growing
(e.g. Refs.~\cite{Mazur:2004fk,Visser:2003ge,Visser:2009pw}
and see Ref.~\cite{Cardoso:2017cqb} for a review).
The exterior of such (static) objects is described by the same Schwarzschild geometry as that of a nonspinning BH.
Thus, as we stressed already, it is challenging to find evidence of a surface using classical 
electromagnetic or gravitational waves~\cite{Abramowicz:2002vt,Chirenti:2007mk,Sakai:2014pga,Kubo:2016ada,Chirenti:2016hzd,Cardoso:2017cqb,Cardoso:2016rao,Cardoso:2017njb,Cardoso:2017cfl,Maselli:2017cmm,Shaikh:2018lcc,Cunha:2018gql,Berthiere:2017tms}.

Classical physics predicts measurable differences between ultracompact horizonless stars and BHs, but these may either be inaccessible
to observers far away or simply take too long to affect our detectors.
However, there is a semiclassical effect which is, seemingly, particular to BH geometries: Hawking radiation.
In fact, when quantum effects are included at a semiclassical level,
particles are created and emitted by BHs, and the spectrum of the
radiation is thermal, such as that of a black body~\cite{Hawking:1974rv,Hawking:1974sw,Birrell:1982ix} . 
In Refs.~\cite{Barcelo:2007yk,Barcelo:2010xk,Kinoshita:2011qs,Chen:2017pkl,Unruh:2018jlu}, quantum particle creation by a collapsing object and its semiclassical effect on the formation of an apparent horizon have been discussed, based on quantum field theory in curved spacetime, in a very general context. Quantum particle creation in horizonless gravitational collapse has
also been discussed in the context of naked singularity
formation~\cite{Ford:1978ip,Hiscock:1982pa,Barve:1998ad,Barve:1998tv,Vaz:1998gd,Harada:2000jya,Harada:2000ar,Harada:2001nj}. 

The organization and summary of this work is the following. In Sec.~\ref{sec:particle_creation}, 
we review quantum particle creation in spherically symmetric spacetimes. 
In Sec.~\ref{sec:spherical_shell}, we expand on our toy model of a collapsing spacetime by pasting Minkowski and Schwarzschild spacetimes with a timelike shell. In Sec.~\ref{sec:standard_collapse}, we review how the present formalism can be used to recover a constant particle radiation by BHs, i.e., the Hawking
radiation, with an emphasis on transient thermal radiation in the
absence of horizon formation.
In Sec.~\ref{sec:null_shell_model}, we introduce a collapse model
with a null shell to a horizonless compact object, yielding
delta-functional divergent emissions both at the end of the transient
Hawking radiation and at the end of the long dormant stage. In Sec.~\ref{sec:timelike_shell_model}, we construct a collapse model
with a timelike shell to a horizonless compact object, 
show the couple of finite bursts of radiation as a common
feature in a broad class of
models, and present the temporal change of radiation for specific models.
Sec.~\ref{sec:discussion} is devoted to discussion. 
We use units in which $G=c=\hbar=1$.

\section{Particle creation in spherically symmetric spacetimes}
\label{sec:particle_creation}
\begin{figure}[htbp]
 \begin{center}
\begin{tabular}{ccc}
\subfigure[Collapse to a static star]{\includegraphics[width=0.3\textwidth]{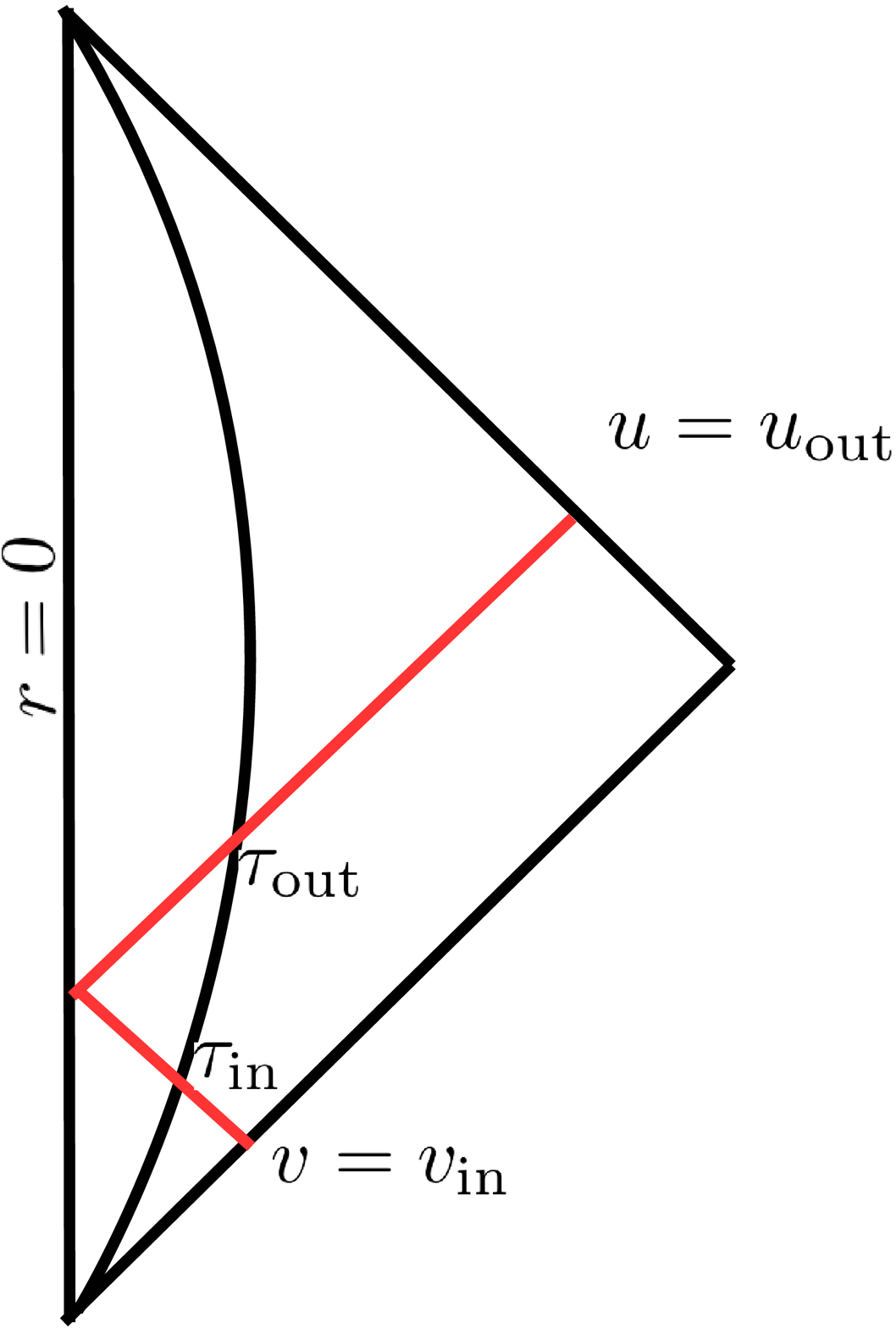}}
 & \quad\quad  &
\subfigure[Collapse to a black hole]{\includegraphics[width=0.3\textwidth]{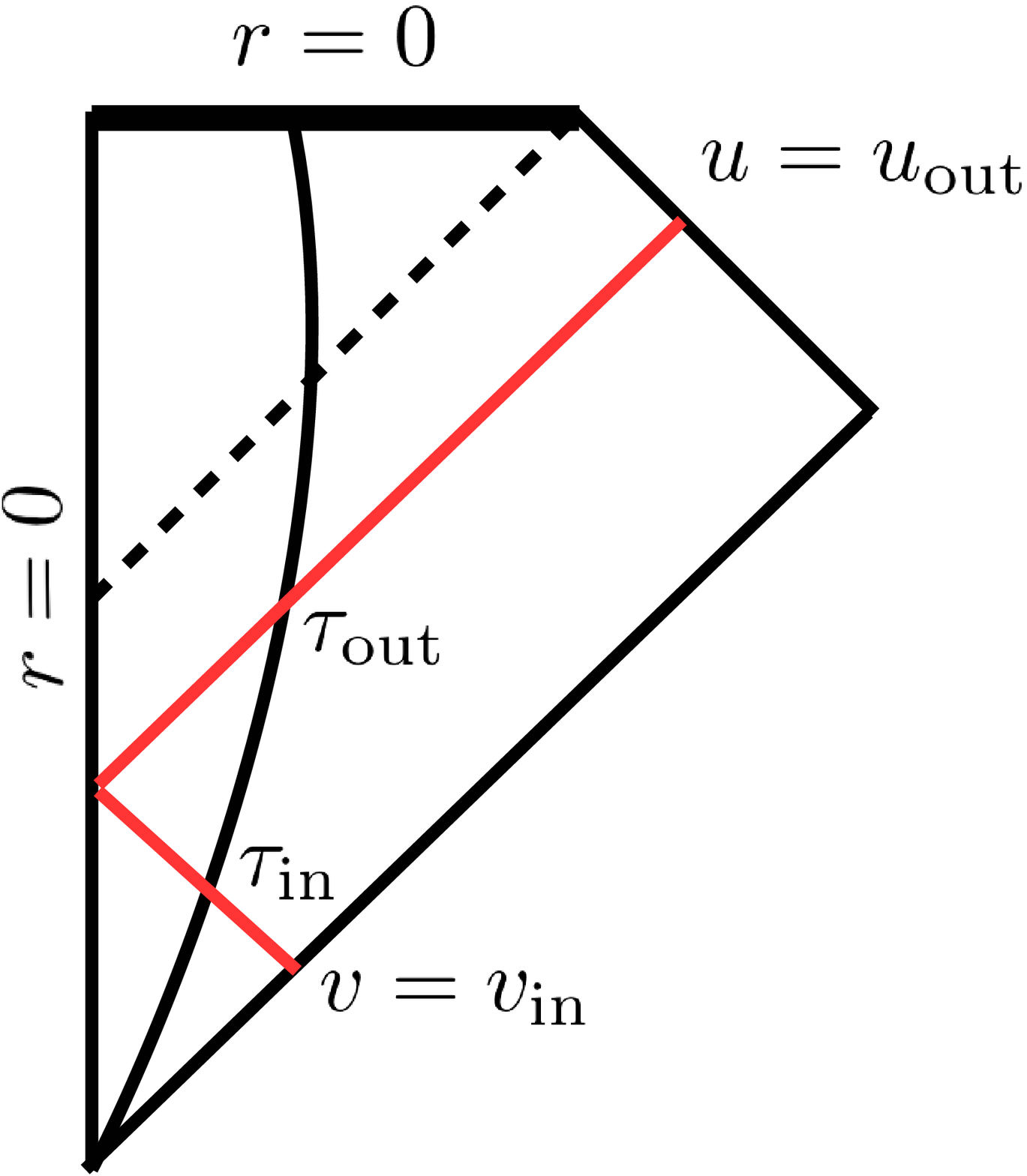}}
\end{tabular}
\caption{
The conformal diagrams for the spacetimes of collapse to (a) a static star and 
(b) a black hole. A pair of outgoing and ingoing null
  rays with $u=u_{\rm out}$ and $v=v_{\rm in}$, respectively,
is also depicted in each diagram. \label{fg:v_G_u}
}
\end{center}
\end{figure}
Consider a spherically symmetric asymptotically flat spacetime.
Let $u$ and $v$ be radial null coordinates, which  
can be written as $u=t-r$ and $v=t+r$ in the asymptotic region, where
$(t,r,\theta,\phi)$ are the usual quasi-Minkowskian spherical 
coordinates there.
Consider a pair of ingoing and outgoing null rays, 
$v=v_{\rm in}$ and $u=u_{\rm out}$,
respectively, which are connected at the regular center $r=0$ with each
other. The null-ray pairs are depicted in 
the conformal diagrams for the spacetimes of collapse to a static star
and a black hole in Figs.~\ref{fg:v_G_u}(a)
and \ref{fg:v_G_u}(b), respectively. 
The mapping function $G$ is defined as $v_{\rm in}=G(u_{\rm out})$.
Note that $u$ can be identified with the observer's time at infinity.
Following Refs.~\cite{Barcelo:2010xk,Kinoshita:2011qs}, we define 
\begin{equation}
\kappa(u_{\rm out}):=-\frac{d}{du_{\rm out}}\ln\frac{dv_{\rm in}}{du_{\rm out}}= -(\ln
G')'(u_{\rm out})\,, \label{eq:kappa}
\end{equation}
which is physically interpreted as the growth rate of redshift of the
outgoing photon with respect to the ingoing photon 
as a function of the retarded
time $u_{\rm out}$. 

To calculate quantum radiation, for simplicity, we adopt the same set of assumptions as in
Refs.~\cite{Hawking:1974rv,Hawking:1974sw,Ford:1978ip}. That is, we assume 
Gaussian (non-self-interacting) massless 
scalar fields, adopt the geometrical optics approximation and 
take a quantum state containing no particle associated with a mode function 
$F_{\omega l m}$ which takes the form $F_{\omega l m}\sim (4\pi \omega)^{-1/2}r^{-1}e^{-i\omega v} Y_{lm}(\theta,\phi)$ at 
past null infinity as an initial quantum state. Then, 
the function $\kappa(u)$ determines radiation power regularized at future null 
infinity, through~\cite{Ford:1978ip,Birrell:1982ix,Harada:2000ar}
\begin{equation}
 P_{lm}=\frac{1}{48\pi}(\kappa^{2}+2\delta \kappa')\, , \label{eq:P_kappa}
\end{equation}
with $\delta=1$ and $0$ for minimally and conformally coupled massless
scalars, respectively, for each $(l,m)$ mode. Note that for higher $l$'s, the geometrical optics
approximation is not valid and the power is strongly suppressed 
due to backscattering. Thus, the total power is dominated by 
sufficiently low $l$'s. 
We can thus omit the $(lm)$ subscript in Eq.~(\ref{eq:P_kappa})
and regard its right-hand side
as an order-of-magnitude estimate of the total power of radiation.
The second term in parentheses on the right-hand side 
of Eq.~(\ref{eq:P_kappa})
does not contribute to the
integrated radiated energy because it is a total derivative; hence, we will mainly concentrate on the first term, i.e., that for the conformally coupled
scalar field. 

If the function $\kappa(u)$ satisfies the adiabatic condition
\begin{equation}
|\kappa'(u_{*})|\ll \kappa^{2}(u_{*})\,,\label{eq:adiabatic_condition}
\end{equation}
then the spectrum of outgoing particles at $u=u_{*}$ can be regarded as Planckian with temperature $T$~\cite{Barcelo:2010xk,Kinoshita:2011qs},
\begin{equation}
kT(u_{*})=\frac{\kappa(u_{*})}{2\pi}\,,
\label{eq:time-dependent_temperature}
\end{equation}
where $\kappa(u_{*})>0$ is assumed.

\section{Spherical shell in vacuum}
\label{sec:spherical_shell}
Our model is a spherically symmetric vacuum spacetime with a shell. The
areal radius of a timelike shell is given by $r=R(\tau)$, where $\tau$
is the proper time for the observer at rest on the shell. 
The induced metric on the timelike world tube $\Sigma$ is given by 
\begin{equation}
ds_{\Sigma}^{2}=-d\tau^{2}+R^{2}(\tau)d\Omega^{2}\, ,
\end{equation}
where $d\Omega^{2}=d\theta^{2}+\sin^{2}\theta d\phi^{2}$ is the metric
on a unit sphere. 
The interior is described by the Minkowski metric
\begin{equation}
 ds^{2}=-dT^{2}+dr^{2}+r^{2}d\Omega^{2}\,.
\end{equation}
The null coordinates in the interior are $U=T-r$ and $V=T+r$.
The exterior is given by the Schwarzschild metric
\begin{equation}
ds^{2}=-\left(1-\frac{2M}{r}\right)dt^{2}+\left(1-\frac{2M}{r}\right)^{-1}dr^{2}+r^{2}d\Omega^{2}\,.
\end{equation}
The standard null coordinates are given by $u=t-r^{*}$ and $v=t+r^{*}$, where $r^{*}:=r+2M\ln \left[({r}/{2M})-1\right]$.
The junction condition for the first fundamental form gives $\dot{t}$
and $\dot{T}$, where the dot denotes the derivative with respect to
$\tau$. This gives $\dot{U}$ and $\dot{V}$ and $\dot{u}$ and
$\dot{v}$. The explicit expressions for them are relegated to Appendix~\ref{sec:ABCD_general}.

Since $V=V_{\rm in}$ and $U=U_{\rm out}$ are related through $V_{\rm
in}=U_{\rm out}$ at the center $r=0$, we find
\begin{eqnarray}
 G'(u)&=&\frac{dv_{\rm in}}{du_{\rm out}}=\frac{dv_{\rm in}}{d\tau_{\rm in}}\frac{d\tau_{\rm in}}{dV_{\rm in}}\frac{dV_{\rm in}}{dU_{\rm out}}\frac{dU_{\rm out}}{d\tau_{\rm out}}\frac{d\tau_{\rm out}}{du_{\rm out}}=\frac{A_{\rm out}}{B_{\rm in}},\label{eq:G'_out_in}
\end{eqnarray}
where 
\begin{eqnarray}
 A:=\frac{\dot{U}}{\dot{u}}~~\mbox{and}~~
 B:=\frac{\dot{V}}{\dot{v}}\,,
\label{eq:A_B_def}
\end{eqnarray}
$A_{\rm out}=A(\tau_{\rm out}(u))$ and so on, and $\tau_{\rm out}(u)$
and $\tau_{\rm in}(u)$ are the values of $\tau$ when the outgoing and
ingoing null rays cross the shell, respectively, as shown in
Figs.~\ref{fg:v_G_u} (a) and (b). 
Further, we can obtain the expression for $\kappa(u)$ as follows:
\begin{equation}
\kappa(u)=C_{\rm out}-\frac{A_{\rm out}}{B_{{\rm in}}}D_{\rm in}\,,\label{eq:kappa_general}
\end{equation}
where
\begin{eqnarray}
 C:=-\frac{1}{\dot{u}}\frac{d\ln A}{d\tau}~~\mbox{and}~~ 
 D:=-\frac{1}{\dot{v}}\frac{d\ln B}{d\tau}\,.
\label{eq:C_D_def}
\end{eqnarray}
The first and second terms on the right-hand side of
Eq.~(\ref{eq:kappa_general}) can be regarded as the contributions from the shell at $\tau=\tau_{\rm out}$
and $\tau=\tau_{\rm in}$, respectively. We can obtain 
the explicit expressions for $A$, $B$, $C$, and $D$ in terms of $R$,
$\dot{R}$, and $\ddot{R}$, which are 
relegated to Appendix~\ref{sec:ABCD_general}.

For reference, if the shell is marginally bound and made of dust, then the junction condition for the second
fundamental form gives
\begin{eqnarray}
\dot{R}^{2}=-1+\left(1+\frac{M}{2R}\right)^{2}\,
~~\mbox{and}~~\ddot{R}&=&-\frac{M}{2R^{2}}\left(1+\frac{M}{2R}\right).
\label{eq:dust_shell_collapse}
\end{eqnarray}
However, we will {\it not} assume any equation of state for the surface energy density and pressure on the shell. Instead, 
we specify the dynamics of the shell. The evolution of the surface energy density and pressure 
will then be determined by the junction condition for the second fundamental form. This freedom has a price: our simplistic model may contain unphysical matter content
with an exotic equation of state. 
We should stress that our purpose here is not to produce alternatives to BHs; rather, we are interested 
in understanding possible consequences of failing to produce horizons. This program, if successful, then allows us to quantify in a better way the evidence for BHs and to strengthen
the BH paradigm.

\section{Particle creation in standard-collapse phase
 \label{sec:standard_collapse}}

Conventionally, to derive the Hawking radiation, the expandability of
$R(\tau)$ with respect to $\tau$ at the entry into the horizon 
$\tau=\tau_{H}$ has often been assumed~\cite{Birrell:1982ix}. 
However, such an assumption seems to prescribe the behavior of the shell at an event which is not in the 
causal past of the observer. Here we show that the expandability at $\tau=\tau_{H}$
is unnecessary and, hence, that the (temporarily) thermal radiation does not 
need any horizon.

Instead, observing the dust-shell collapse described by 
Eq.~(\ref{eq:dust_shell_collapse}),
we assume that the standard collapse is divided into the following two phases:
\begin{itemize}
 \item[(i)] Phase 0, an early-collapse phase: $\tau<\tau_{0}$ or $R>4M$.\\
We assume
\begin{equation}
1-\frac{2M}{R}> \frac{1}{2}, \quad |\dot{R}|\lesssim 1, \quad
 \mbox{and}\quad 
 |\ddot{R}|\lesssim \frac{1}{2M}.
\end{equation}
We can additionally assume that the shell is initially static at some radius
       $R_{i}$.
 \item[(ii)] Phase 1, a late-collapse phase: $\tau>\tau_{0}$ or $2M<R<4M$.\\
We assume
\begin{equation}
 1-\frac{2M}{R}< \frac{1}{2},~~ 1-\frac{2M}{R}< \dot{R}^{2},~~\dot{R}=O(1),~~ \mbox{and}~~ \ddot{R}=O((2M)^{-1}).
\end{equation}
\end{itemize}
The functions $A$, $B$, $C$, and $D$ take expressions $A_{j}$, $B_{j}$,
$C_{j}$, and $D_{j}$ for phase $j$. 
The explicit expressions are relegated to Appendix \ref{sec:ABCD}.
The transition between the above two regimes occurs at $\tau=\tau_{0}$
when $R= 4M$. This scenario of standard collapse is then consistent with the dust-shell collapse.

Denoting the outgoing null ray in the Schwarzschild region 
which leaves the shell at $\tau=\tau_{0}$ with $u=u_{0}$, 
we can obtain the expression for $G'(u)$ and $\kappa(u)$
separately for $u<u_{0}$ and $u>u_{0}$.
To do this, it is a key to determine when the outgoing null ray crosses
the shell outwardly and when the ingoing null ray, which is a
counterpart of the outgoing null ray in the pair, crosses the shell
inwardly.
If the outgoing null ray crosses the shell outwardly in phase $i$ at
$\tau=\tau_{\rm out}$ and  the ingoing null ray crosses the shell
inwardly in phase $j$ at $\tau=\tau_{\rm in}$, 
we classify the null-ray pair as $(i,j)$.
For the null-ray pair of class $(i,j)$, $G'$ and $\kappa$ are given by 
\begin{eqnarray}
 G'(u)=\frac{A_{i,{\rm out}}(u)}{B_{j,{\rm in}}(u)} ~~\mbox{and}~~
 \kappa(u)=C_{i,{\rm out}}(u)-\frac{A_{i,{\rm out}}(u)}{B_{j,{\rm in}}(u)}D_{j,{\rm in}}(u),
\label{eq:general_formula_for_G'_kappa}
\end{eqnarray}
respectively, where we use the notation $A_{i,{\rm out}}(u)=A_{i}(\tau_{{\rm out}}(u))$ and so on.
Then, we can find that there are two radiation stages.
\begin{itemize}
\item[(i)] $u<u_{0}$\\
Since $\tau_{\rm in}<\tau_{\rm out}<\tau_{0}$, 
the null-ray pairs are classified as $(0,0)$. 
Using Eqs.~(\ref{eq:general_formula_for_G'_kappa}) and 
(\ref{eq:1_ABCD}), we have 
\begin{eqnarray}
\kappa(u)\simeq   -\left[\frac{M}{R}\left(\ddot{R}-\frac{|\dot{R}|}{R}\right)\right]_{\rm out}
-\left[\frac{M}{R}\left(\ddot{R}+\frac{|\dot{R}|}{R}\right)\right]_{\rm in}\,.
\end{eqnarray}
Therefore, we can conclude that $|\kappa|\lesssim 1/(4M)$.
Thus, the radiation for $u<u_{0}$, which may be called pre-Hawking radiation, 
is weaker than the standard Hawking radiation.

 \item[(ii)] $u>u_{0}$\\
For $\tau_{\rm in}<\tau_{0}$, the null-ray pairs are classified as
$(1,0)$, while,
for $\tau_{\rm in}>\tau_{0}$, the null-ray pairs are classified as $(1,1)$.
For both cases, from Eqs.~(\ref{eq:general_formula_for_G'_kappa}), 
(\ref{eq:1_ABCD}), and (\ref{eq:late-collapse_phase})
we have the same expression for $\kappa(u)$:
\begin{equation}
\kappa(u)\simeq  C_{\rm out} \simeq \frac{1}{4M}\,.\label{eq:kappa_standard_collapse}
\end{equation}
\end{itemize}
Here, we discuss the radiation for $u>u_{0}$.
Using Eq.~(\ref{eq:P_kappa}), we obtain
\begin{equation}
P\simeq P_{H}=\frac{1}{48\pi}\frac{1}{16M^{2}}\,.\label{eq:Hawking_radiation_power}
\end{equation}
This is the reproduction of the Hawking radiation.
Since the first term is dominant in the expression for $\kappa(u)$ in
Eq.~(\ref{eq:general_formula_for_G'_kappa}), the Hawking radiation (whether transient or eternal) originates from the behavior of
the shell in the late-collapse phase at $\tau=\tau_{\rm out}$.

Equations~(\ref{eq:kappa}) and (\ref{eq:time-dependent_temperature})
give temporarily thermal radiation with temperature
\begin{equation}
kT(u_{*})\simeq kT_{H}=\frac{1}{8\pi M}\,,
\end{equation}
where we can easily see that the adiabatic condition
(\ref{eq:adiabatic_condition}) is also satisfied.
Since no horizon has formed yet, this means that transient Hawking radiation does not need any horizon.
If the late-collapse phase continues up until $R\simeq 2M(1+\epsilon^{2})$, then the transient Hawking radiation arises and lasts for $\Delta u \simeq 4M \ln \epsilon^{-2}$, which can be seen from 
Eq.~(\ref{eq:u_v_R_late-collapse}). Therefore, 
the radiated energy through this transient
Hawking radiation is given by 
\begin{equation}
 E\simeq \frac{1}{48\pi }\frac{\ln\epsilon^{-2}}{4M}.
\end{equation}
In the limit $\epsilon\to 0$, which may correspond to the formation of 
an event horizon depicted in Fig.~\ref{fg:v_G_u}(b), 
the Hawking radiation continues eternally
and the energy radiated goes to infinity.

\section{Collapse to an ultracompact object with a null shell }
\label{sec:null_shell_model}
%
\begin{figure}[htbp]
\begin{center}
\includegraphics[height=0.4\textheight]{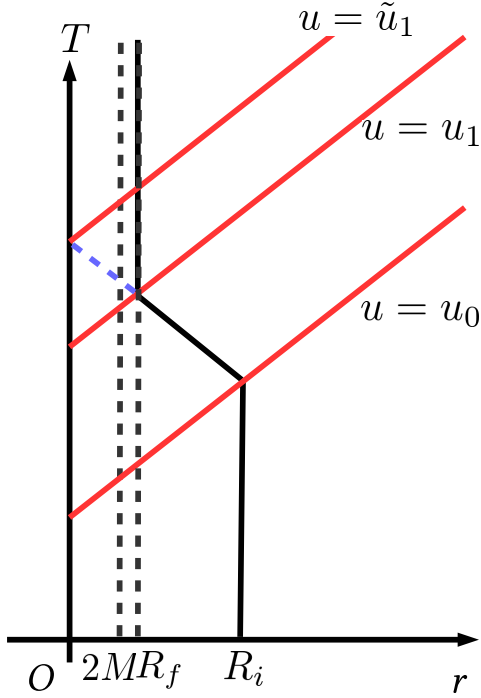}
\caption{\label{fg:null_shell_model} The collapse model with an ingoing null shell. The static shell at $R=R_{i}$ changes to an ingoing null shell at $u=u_{0}$ and again becomes static with
$R=R_{f}$ at $u=u_{1}$. 
The ingoing null shell is extended to the Minkowski region with an
 ingoing null ray, which is denoted by a blue dashed line, and reflected
 to an outgoing null ray that passes the shell outwardly 
to the Schwarzschild region, which is denoted with a red line labeled $u=\tilde{u}_{1}$. This model was introduced in Ref.~\cite{Paranjape:2009ib}.}
 \end{center}
\end{figure}
We now review and reanalyze an exact collapse model with a null shell, 
which can result in a static compact star with radius slightly 
larger than $2M$. 
The schematic diagram of this model---introduced in
Ref.~\cite{Paranjape:2009ib}---is shown in
Fig.~\ref{fg:null_shell_model} and consists of three phases.
Note that these phases are different from those in the timelike-shell model.
\begin{itemize}
 \item[(i)] $u<u_{0}$\\
Initially, the shell is static with $R=R_{i}$.
 \item[(ii)] $u_{0}<u<u_{1}$\\ 
At $u=u_{0}$, the shell suddenly 
turns ingoing null with $V=0$. Since the shell 
is also given by $v=$const., we find 
\begin{eqnarray}
u-u_{0}=U-4M\ln\left(\frac{-U}{4M}-1\right)+2R_{i}+4M\ln\left(\frac{R_{i}}{2M}-1\right)\,,
\label{eq:u_U_null_shell}
\end{eqnarray}
where $U=-2R_{{\rm out}}$ is a monotonically increasing function of $u$ from $-2R_{i}$ to $-2R_{f}$.
\item[(iii)] $u>u_{1}$\\
When the shell reaches the radius $R_{f}:=2M/(1-\epsilon^{2})$ at $u=u_{1}$, 
it stops and becomes static again, where $u_{1}$ is determined by 
\begin{equation}
 u_{1}-u_{0}=-2\left[\frac{2M}{1-\epsilon^{2}}+2M\ln\frac{\epsilon^{2}}{1-\epsilon^{2}}\right]
+2\left[R_{i}+2M\ln\left(\frac{R_{i}}{2M}-1\right)\right].
\end{equation}
We treat $\epsilon$ as a constant free parameter satisfying $0<\epsilon< 1$.
\end{itemize}
We also define $\tilde{u}_{1}$ such that the ingoing null shell $V=0$ is
extended with an ingoing null ray to the center $r=0$ in the Minkowski
region, being reflected to the outgoing null ray and going through the shell to an outgoing null
ray $u=\tilde{u}_{1}$ in the Schwarzschild region. We can find
\begin{equation}
 \tilde{u}_{1}=u_{1}+\frac{4M}{\epsilon(1-\epsilon^{2})}\,.
\end{equation}
The functions $G'(u)$ and $\kappa(u)$ are calculated as follows:
\begin{itemize}
\item[(i)] $u<u_{0}$\\
All null-ray pairs are classified as $(0,0)$, for which 
we have $G'(u)=1$ and $\kappa(u)=0$.
\item[(ii)] $u_{0}<u<u_{1}$\\ 
All null-ray pairs are classified as $(1,0)$, for which 
\begin{eqnarray}
 G'(u)=\left(1-\frac{2M}{R_{i}}\right)^{-1/2}\left(1+\frac{4M}{U}\right),\quad 
 \kappa(u)=-\frac{G''}{G'}=\frac{4M}{U^{2}}, 
\end{eqnarray}
where $U(u)$ is implicitly given by Eq.~(\ref{eq:u_U_null_shell}).
\item[(iii)] $u_{1}<u<\tilde{u}_{1}$\\ 
All null-ray pairs are classified as $(2,0)$, for which  
\begin{equation}
G'(u)=\epsilon\left(1-\frac{2M}{R_{i}}\right)^{-1/2},\quad \kappa(u)=0.
\end{equation}
\item[(iv)] $u>\tilde{u}_{1}$\\
All null-ray pairs are classified as $(2,2)$, for which we have
$G'(u)=1$ and $\kappa(u)=0$.
\end{itemize}
Therefore, particles are emitted for $u_{0}<u<u_{1}$, but not for $u<u_{0}$,
$u_{1}<u<\tilde{u}_{1}$ and $\tilde{u}_{1}<u$. 
For $\epsilon\ll 1$, the radiation for
$u_{0}<u<u_{1}$ can be regarded as temporarily thermal with temperature
$kT=\kappa(u)/(2\pi)=M/(2\pi R_{{\rm out}}^{2})$. Therefore, $kT\simeq
1/(8\pi M)$ for $1-2M/R_{{\rm out}}\ll 1$. This is transient Hawking radiation.

\begin{figure}[htbp]
\begin{center}
\includegraphics[height=0.3\textheight]{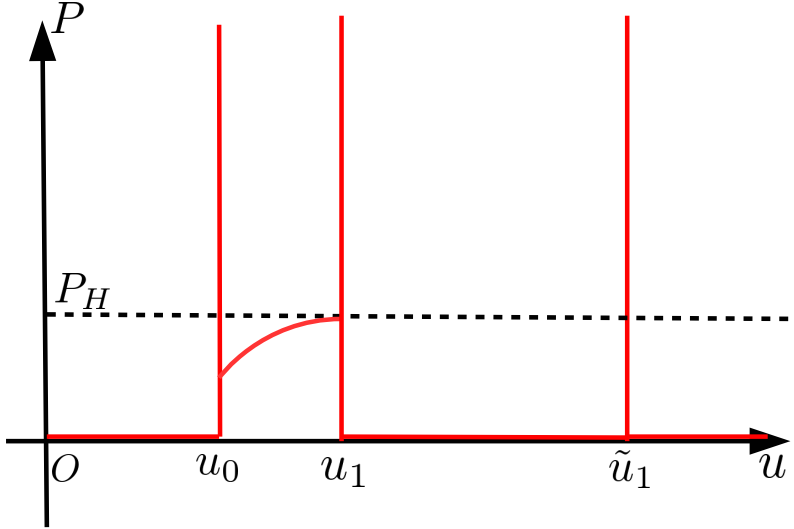}
\caption{\label{fg:power_evolution_null_shell} The schematic figure of
 the evolution of the power in the collapse to a highly compact object
 in the null-shell model introduced in Ref.~\cite{Paranjape:2009ib}. The three vertical lines denote
delta-functional divergences at $u=u_{0}, u_{1}$ and $\tilde{u}_{1}$,
 while there appears transient Hawking radiation for $u_{0}<u<u_{1}$.}
\end{center}
\end{figure}
In this model, we have bursts of radiation 
at $u=u_{0}$, $u_{1}$, and $\tilde{u}_{1}$ because $G'$ changes
discontinuously then and $\kappa$ is given by Eq.~(\ref{eq:kappa}). 
The discontinuities in $(-\ln G')$, which we denote with $\Delta(-\ln
G')$, 
are given as follows:
\begin{eqnarray}
 \Delta(-\ln G')_{u=u_{0}}&=&\ln\left(1-\frac{2M}{R_{i}}\right)^{-1/2}, \\
 \Delta(-\ln G')_{u=u_{1}}&=& -\ln\epsilon^{-1}, \\ 
 \Delta(-\ln G')_{u=\tilde{u}_{1}}&=&-\ln\epsilon^{-1}+\ln\left(1-\frac{2M}{R_{i}}\right)^{1/2}.
\end{eqnarray}
More precisely, the bursts are described by the square of a delta
function at $u=u_{0}, u_{1}$ and $\tilde{u}_{1}$, which suggests
infinite radiated energy in an infinitesimal span of time.
We schematically plot the evolution of the power of radiation in Fig.~\ref{fg:power_evolution_null_shell}.
The discontinuity in $(-\ln G')$ is positive and $O(1)$ at $u=u_{0}$,
while it is negative and $O(\ln \epsilon^{-1})$ at both $u=u_{1}$ and
$u=\tilde{u}_{1}$ for $\epsilon\ll 1$.
[This divergent behavior was overlooked in Ref.~\cite{Paranjape:2009ib}. For example, in Fig.~2 of Ref.~\cite{Paranjape:2009ib}, 
there should be three vertical lines indicating delta-functional
divergences at $u/(2M)=u_{0}/(2M)=0$, $u_{1}/(2M)\simeq 19.49$, and
$\tilde{u}_{1}/(2M)\simeq 29.91$
for the choice $u_{0}=0$, $R_{i}=12M$ and $\epsilon=0.2$.]
The delta-functional burst at $u=u_{0}$ can be removed if the onset 
of the collapse process is adiabatic.
On the other hand, the bursts 
at $u=u_{1}$ and $\tilde{u}_{1}$ are of more physical interest.

\section{Collapse to an ultracompact object with a timelike shell}
\label{sec:timelike_shell_model}

\subsection{Phases of the shell dynamics}
The features discussed in the null-shell model (in particular the bursts
of radiation both at the end of transient Hawking radiation and at the
end of the long dormant stage) are of physical interest. However, the
delta-functional divergence is clearly unphysical and arises from the
instantaneous transitions from the static shell to null at $u=u_{0}$ and
the null shell to timelike at $u=u_{1}$. We also see that the power emitted is finite, as
long as $\ddot{R}$ and $\dot{R}$ are finite
[cf. Eq.~\eqref{eq:kappa_general}
and Appendix~\ref{sec:ABCD_general}].

\begin{figure}[htbp]
 \begin{center}
  \includegraphics[height=0.5\textheight]{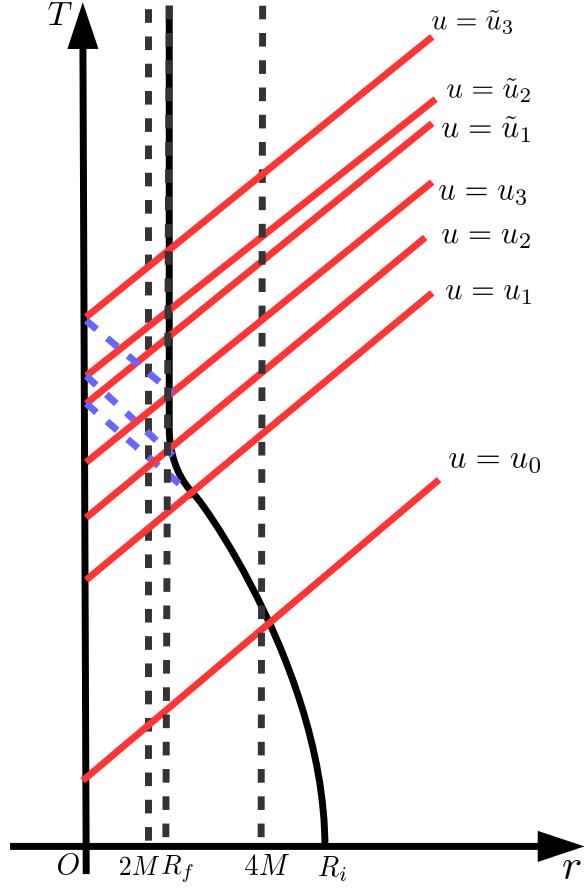}
\caption{\label{fg:timelike_shell_model}
The collapse model with a timelike shell. The shell enters $R=4M$ at
  $\tau=\tau_{0}$, begins to brake at $\tau=\tau_{1}$, and stops at
  $\tau=\tau_{3}$. Between $\tau_{1}$ and $\tau_{3}$, there is a moment
  $\tau_{2}$, when the equality $1-2M/R=\dot{R}^{2}$ is satisfied. 
The outgoing null rays which
pass the shell outwardly to the Schwarzschild region at $\tau=\tau_{0}$,
  $\tau_{1}$, $\tau_{2}$, and $\tau_{3}$ are denoted by red lines
  labeled $u=u_{0}$, $u_{1}$, $u_{2}$ and $u_{3}$, 
respectively. The ingoing null rays denoted by blue dashed lines
leave the shell at $\tau=\tau_{1}$, $\tau_{2}$ and $\tau_{3}$, 
and are reflected to outgoing null rays passing the shell outwardly 
to the Schwarzschild region denoted by 
red lines labeled $u=\tilde{u}_{1}$, $\tilde{u}_{2}$ and
  $\tilde{u}_{3}$, respectively.
}
 \end{center}
\end{figure}
To have a smooth process and extract meaningful physics, we propose a
collapse model of a timelike shell with finite $\ddot{R}$ and $\dot{R}$.
This model consists of five phases: an early-collapse phase,
late-collapse phase, early-braking phase, late-braking phase, and 
final static phase.
\begin{itemize}
\item Phase 0, an early-collapse phase: $\tau<\tau_{0}$ or $R>4M$.\\ 
This phase is identical to that in standard collapse discussed in Sec.~\ref{sec:standard_collapse}; in particular we assume
that $1-2M/R> 1/2$, $|\dot{R}|\lesssim 1$, and $|\ddot{R}|\lesssim 1/(2M)$.

\item Phase 1, a late-collapse phase: $\tau_{0}<\tau<\tau_{1}$ or $R_{b}<R<4M$.\\
This phase is also identical to that in standard collapse discussed in Sec.~\ref{sec:standard_collapse},
i.e., $1-2M/R< 1/2$, $1-2M/R< \dot{R}^{2}$, $\dot{R}=O(1)$, and $\ddot{R}=O((2M)^{-1})$.

\item Phase 2, an early-braking phase: $\tau_{1}<\tau<\tau_{2}$ or $R_2<R<R_b$.\\ 
We assume that at $\tau=\tau_{1}$ or $R=R_b$, the shell begins to brake.
For $\tau_{1}<\tau<\tau_{2}$, we assume the following inequality:
\begin{eqnarray}
1-\frac{2M}{R}< \dot{R}^{2}\, .
\label{eq:hierarchy}
\end{eqnarray}

\item Phase 3, a late-braking phase: $\tau_{2}<\tau<\tau_{3}$ or $R_f< R<R_2$\\
We assume that at $\tau=\tau_{2}$, when $R=R_{2}$, the following
      equality holds: 
\begin{eqnarray}
1-\frac{2M}{R}= \dot{R}^{2}\, .
\label{eq:hierarchy_marginal}
\end{eqnarray}
For $\tau_{2}<\tau<\tau_{3}$, the following inequality holds: 
\begin{eqnarray}
1-\frac{2M}{R}> \dot{R}^{2}\, .
\label{eq:hierarchy_opposite}
\end{eqnarray}
The radius of the shell approaches the final value $R_{f}$.

\item Phase 4, a final static phase: $\tau>\tau_{3}$ or $R=R_f$.\\
We assume that the shell smoothly stops at $\tau=\tau_{3}$, when $R=R_{f}=2M/(1-\epsilon^{2})$. Later on, 
the shell is completely static. 
\end{itemize}

For later convenience, 
as is seen in Fig.~\ref{fg:timelike_shell_model}, we label as $u=u_{0}$,
$u_{1}$, $u_{2}$, and $u_{3}$
those outgoing null rays in the Schwarzschild region which leave the shell
outwardly at $\tau=\tau_{0}$, $\tau_{1}$, $\tau_{2}$, and
$\tau_{3}$, respectively. We use $u=\tilde{u}_{1}$, $\tilde{u}_{2}$,
$\tilde{u}_{3}$ for those outgoing null rays which are traced back
through the center to
ingoing null rays and reach the shell at
$\tau=\tau_{1}$, $\tau_{2}$, $\tau_{3}$, respectively.
We denote that the proper times when the outgoing null rays
$u=\tilde{u}_{1}, \tilde{u}_{2}, \tilde{u}_{3}$ cross the shell
outwardly as 
$\tau=\tilde{\tau}_{1}, \tilde{\tau}_{2}, \tilde{\tau}_{3}$, respectively.

\subsection{Post-Hawking burst \label{subsec:post-Hawking_burst}}
We find that the emission of bursts of radiation both at the end of the
transient Hawking radiation and at the end of a long dormant stage
is a general feature of quantum particle creation in setups 
leading to a compact horizonless
object. Here, we briefly describe this phenomenon.

For $u_{1}<u<u_{3}$, the observer receives the 
outgoing null ray which left the shell outwardly in the braking
phase, and which can be traced back to the ingoing null ray which crosses
the shell inwardly in the standard-collapse phase. From
Appendix~\ref{sec:ABCD}, $\kappa(u)$ is estimated as 
\begin{eqnarray}
 \kappa(u)&=&C_{2,out}+O(\epsilon (2M)^{-1})\simeq
  -\left[\frac{\ddot{R}}{2\dot{R}^{2}}\left(1-\frac{2M}{R}\right)\left(1-\frac{|\dot{R}|}{\sqrt{1+\dot{R}^{2}}}\right)\right]_{\rm
  out}+\frac{1}{4M},
\label{eq:kappa_post-Hawking_burst_2} \\
 \kappa(u)&=&C_{3,out}+O(\epsilon (2M)^{-1})\simeq
  -\ddot{R}_{\rm
  out}+\left[\frac{|\dot{R}|}{4M\sqrt{1-\frac{2M}{R}}}\right]_{\rm out},
\label{eq:kappa_post-Hawking_burst_3} 
\end{eqnarray}
for $u_{1}<u<u_{2}$ and $u_{2}<u<u_{3}$, respectively.
Note that the factor $(1-\frac{2M}{R})/\dot{R}^{2}$ is generally  
an increasing function for $u_{1}<u < u_{3}$, which is much smaller than
unity at $u=u_{1}$, unity at $u=u_{2}$, and diverging at $u=u_{3}$. 
In the above expressions, the second term can be regarded as 
the transient Hawking radiation, which keeps constant 
for $u_{1}<u<u_{2}$ and decays for $u_{2}<u<u_{3}$. 
This implies that $u_{2}$ (or $\tau_{2}$) plays a clear physical 
role: it triggers the decay of the transient Hawking radiation.
On the other hand, the first term is negative and  
 dominates the second term if $\ddot{R}\gtrsim 1/(2M)$ for 
 $u_{2}\lesssim  u < u_{3}$. 
 The emission due to the first term completely ends at $u=u_{3}$.
 This gives a burst of radiation at the end of the
 transient Hawking radiation around 
 $u=u_{2}$, which we call a post-Hawking burst. 
 This particle creation occurs due to the braking of the shell 
 at $\tau=\tau_{\rm out}$. The details of the burst depend
 on the specific behavior of the shell in the braking phase.

\subsection{Late-time burst from a static star
  \label{subsec:late-time_burst}}

Next, we consider the interval $\tilde{u}_{1}<u<\tilde{u}_{3}$, when 
the ingoing null ray crosses the shell inwardly in the braking phase and 
the outgoing null ray crosses the shell outwardly in the final
static phase.
In this case, from Appendix~\ref{sec:ABCD}, 
$\kappa$ is negative and estimated as
\begin{eqnarray}
 \kappa(u)&=&-\epsilon\frac{D_{2,{\rm in}}}{B_{2,{\rm in}}}
\simeq -\epsilon\left[\frac{\ddot{R}}{|\dot{R}|\sqrt{1+\dot{R}^{2}}}
\right]_{{\rm in}} 
  -\epsilon\left[\frac{1}{2|\dot{R}|(\sqrt{1+\dot{R}^{2}}-|\dot{R}|)}\right]_{\rm
  in}\frac{1}{4M},
\label{eq:kappa_late-time_burst} \\
 \kappa(u)&=&-\epsilon \frac{D_{3,{\rm in}}}{B_{3,{\rm in}}}
\simeq -\epsilon \left[\frac{\ddot{R}}{\sqrt{1-\frac{2M}{R}}}
\right]_{{\rm in}}
 -\epsilon\left[\frac{|\dot{R}|}{1-\frac{2M}{R}}\right]_{\rm in}\frac{1}{4M},
\end{eqnarray}
for $\tilde{u}_{1}<u<\tilde{u}_{2}$ and $\tilde{u}_{2}<u<\tilde{u}_{3}$, 
respectively.
Therefore, 
if $\ddot{R}\gtrsim 1/(2M)$ at $\tau=\tau_{2}$ in the braking phase, 
the first term in the above expressions dominates $\kappa(u)$ at
$u=\tilde{u}_{2}$ and,
hence, particle creation occurs due to the braking at $\tau=\tau_{{\rm
in}}$.
This may be regarded as the ingoing part of the
post-Hawking burst propagating through the center and becoming an outgoing flux.
Even if $\ddot{R}$ is totally negligible in the braking phase, 
the second term in the above expressions describe
a burst of radiation with a peak $\kappa\simeq -1/(4M)$ at $u=\tilde{u}_{2}$.
This may be regarded as the ingoing part of the transient 
Hawking radiation propagating through the center and becoming an outgoing flux.

Whether the deceleration is effective in particle creation or not, 
the observation of the burst is delayed from the direct
observation of the deceleration at $u=u_{2}$ by 
$\tilde{u}_{2}-u_{2}\simeq 4M/\epsilon$. 

\subsection{Time dependence of particle
  creation\label{subsec:specific_models}}

We will discuss the whole temporal change of radiation for specific
models below.
\subsubsection{Model A: Exponentially slowed-down model}
First, we assume that $R-R_{f}\propto e^{-\sigma\tau}$ for
$\tau_{1}<\tau <\tau_{3}'$ except for the short interval
$\tau'_{3}<\tau<\tau_{3}$, when $R$ smoothly settles 
down to the final fixed radius $R_{f}$ at $\tau=\tau_{3}$, 
by introducing the deceleration parameter $\sigma$ such that
$\ddot{R}= \sigma |\dot{R}|= \sigma^{2}(R-R_{f})$ with 
\begin{equation}
 \sigma= \frac{|\dot{R}_{b}|}{R_{b}-R_{f}},
\label{eq:sigma}
\end{equation}
where $|\dot{R}_{b}|=O(1)$.
We parametrize $R_{b}$ through $R_{b}-R_{f}=2M\epsilon^{2\beta}$. 
For $\beta=1/2$, we have 
\begin{equation}
 1-\frac{2M}{R}\simeq \left\{\begin{array}{cc}
  \epsilon e^{-\sigma(\tau-\tau_{1})} & (\tau_{1}<\tau<\tau_{2}) \\
  \epsilon^{2} & (\tau_{2}<\tau<\tau_{3})
		      \end{array}\right. ,
\end{equation}
while for $\beta\ge 1$, we have 
\begin{equation}
 1-\frac{2M}{R}\simeq \epsilon^{2}
\end{equation}
for $\tau_{1}<\tau<\tau_{3}$.
We assume that $\tau_{3}-\tau_{2}\simeq \sigma^{-1}$ for simplicity. See
Appendix~\ref{sec:duration} for the estimate of $\tau_{2}$.

Assuming $\beta\ge 1/2$ for simplicity, for the post-Hawking burst, 
$\kappa(u)$ peaks at $u=u_{2}$ with
\begin{equation}
\kappa \simeq -\epsilon \sigma \simeq -\frac{1}{2M\epsilon^{2\beta-1}}.
\end{equation}
The peak power and energy radiated in the post-Hawking burst are roughly estimated to
\begin{eqnarray}
 P\simeq \epsilon^{-2(2\beta-1)}P_{H}~~\mbox{and}~~
 E\simeq \epsilon\sigma \sim \frac{1}{2M\epsilon^{2\beta-1}}.
\end{eqnarray}
So, if $\beta>1/2$, the power and the energy radiated in the
post-Hawking burst dominate those of the transient Hawking radiation.

\begin{figure}[htbp]
\begin{center}
\begin{tabular}{cc}
\subfigure[$\beta=1/2$]{\includegraphics[width=0.4\textwidth]{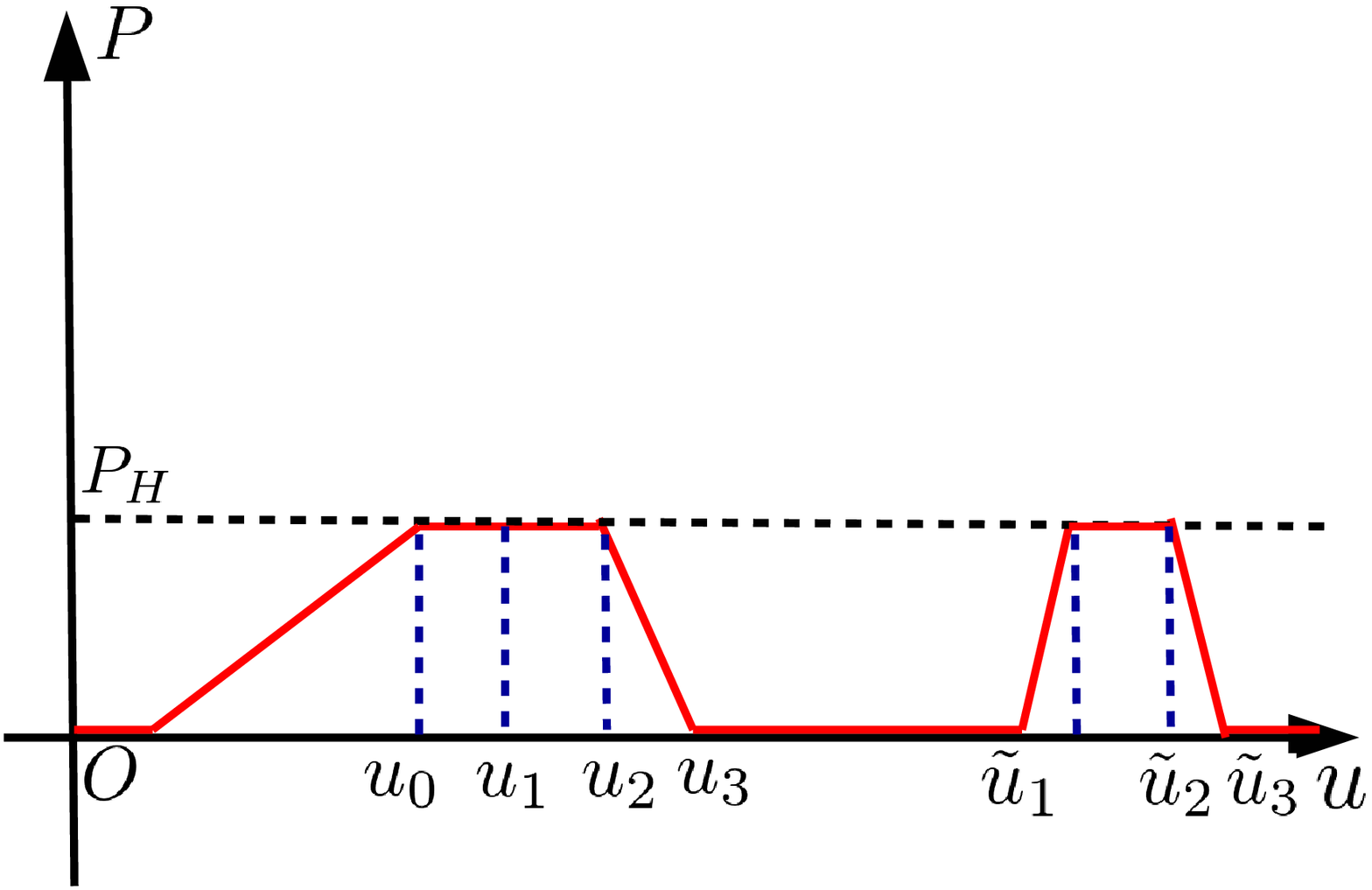}}
& 
\subfigure[
$\beta=1$]{\includegraphics[width=0.4\textwidth]{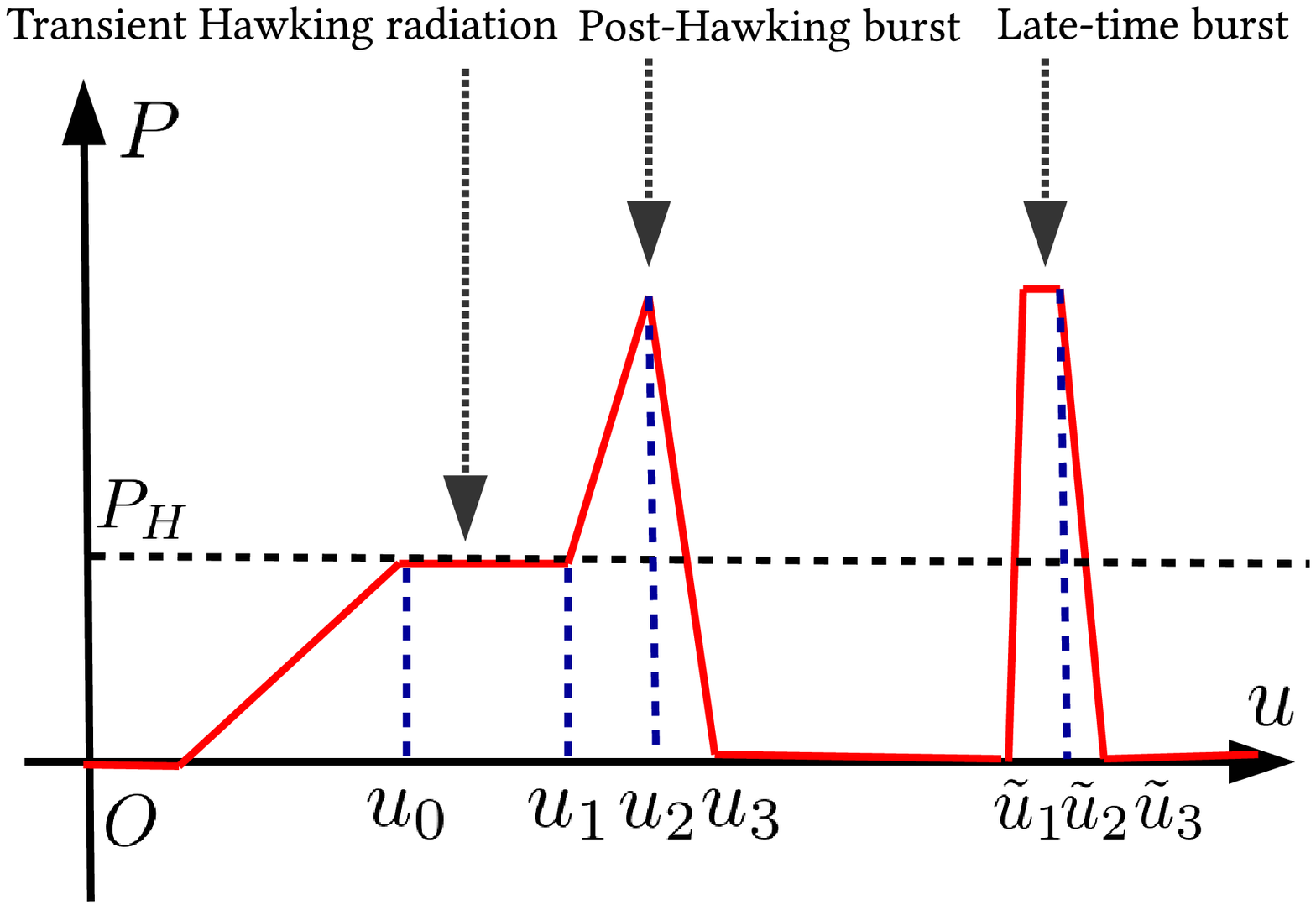}} 
\end{tabular}
\caption{\label{fg:power_evolution} The schematic time dependence of
 radiation emitted in the timelike collapse to a highly compact object
 in model A, in which the shell is exponentially slowed down in the 
 braking phase.
The shell begins braking at $R=R_{b}=R_{f}+2M\epsilon^{2\beta}$ for
 (a) $\beta=1/2$ and (b) $\beta=1$. We here neglect the power of the
 order of $\epsilon^{2}P_{H}$. 
}
\end{center}
\end{figure}

It is interesting to look into the late-time burst.
We can find 
$\kappa(u)$ is nearly constant with
\begin{equation}
 \kappa(u)\simeq -\epsilon\sigma\simeq -\frac{1}{2M\epsilon^{2\beta-1}}
\end{equation}
for $\tilde{u}_{1}<\tilde{u}<\tilde{u}_{2}$ and decays to zero 
for $\tilde{u}_{2}<u<\tilde{u}_{3}$.
The peak power 
is as strong as the Hawking radiation for $\beta=1/2$ 
and is stronger for $\beta>1/2$.
For $\tilde{u}_{1}<u<\tilde{u}_{2}$, we can see
\begin{equation}
|\kappa'|\sim \frac{\epsilon\sigma}{\sigma^{-1}\epsilon^{-1}(\ln\epsilon^{-1})} \sim
 \frac{\kappa^{2}}{\ln \epsilon^{-1}}\ll \kappa^{2}\, ,
\end{equation}
where we have used Eq.~\eqref{eq:tilde_u_2-tilde_u_1}.
Therefore, the late-time burst can be regarded as adiabatic.
However, 
this cannot be interpreted as a Planckian distribution with
negative temperature: the stationary phase approximation or saddle point
approximation to derive the Planck
distribution~\cite{Barcelo:2010xk,Kinoshita:2011qs} is simply not
applicable.~\footnote{The authors are grateful to S.~Kinoshita for
highlighting this issue.} 
The radiated energy during the burst is calculated to 
\begin{equation}
 E\simeq \frac{1}{48\pi }\sigma\epsilon \ln \epsilon^{-1}\simeq
  \frac{1}{48\pi}\frac{\ln \epsilon^{-1}}{2M\epsilon^{2\beta-1}}.
\end{equation}
For $\beta=1/2$,
this is approximately equal to energy radiated through the transient
Hawking radiation, while for $\beta>1/2$, 
this dominates the latter.
The temporal dependences of particle emission are summarized 
for $\beta=1/2$ and $\beta=1$ in Fig.~\ref{fg:power_evolution}. 

\subsubsection{Model B: constant-deceleration model}
Next we consider a technically simpler model, where the deceleration $a$ of the
shell is constant for $\tau_{1}<\tau<\tau_{3}$ with
\begin{equation}
a=\frac{\dot{R}_{b}^{2}}{2(R_{b}-R_{f})},
\label{eq:a}
\end{equation}
where $|\dot{R}_{b}|=O(1)$.
We can naturally assume $a\gg
1/(4M)$. 
Therefore, $\tilde{u}_{3}-\tilde{u}_{1}\sim \epsilon^{-1}a^{-1}$ and 
$\tilde{u}_{3}-\tilde{u}_{2}\sim a^{-1}$ as derived in
Appendix~\ref{sec:duration}. We parametrize
$R_{b}-R_{f}=2M\epsilon^{2\beta}$ ($\beta>0$) as in the previous model.

\begin{figure}[htbp]
\begin{center}
\begin{tabular}{cc}
\subfigure[$\beta=1/2$]{\includegraphics[width=0.4\textwidth]{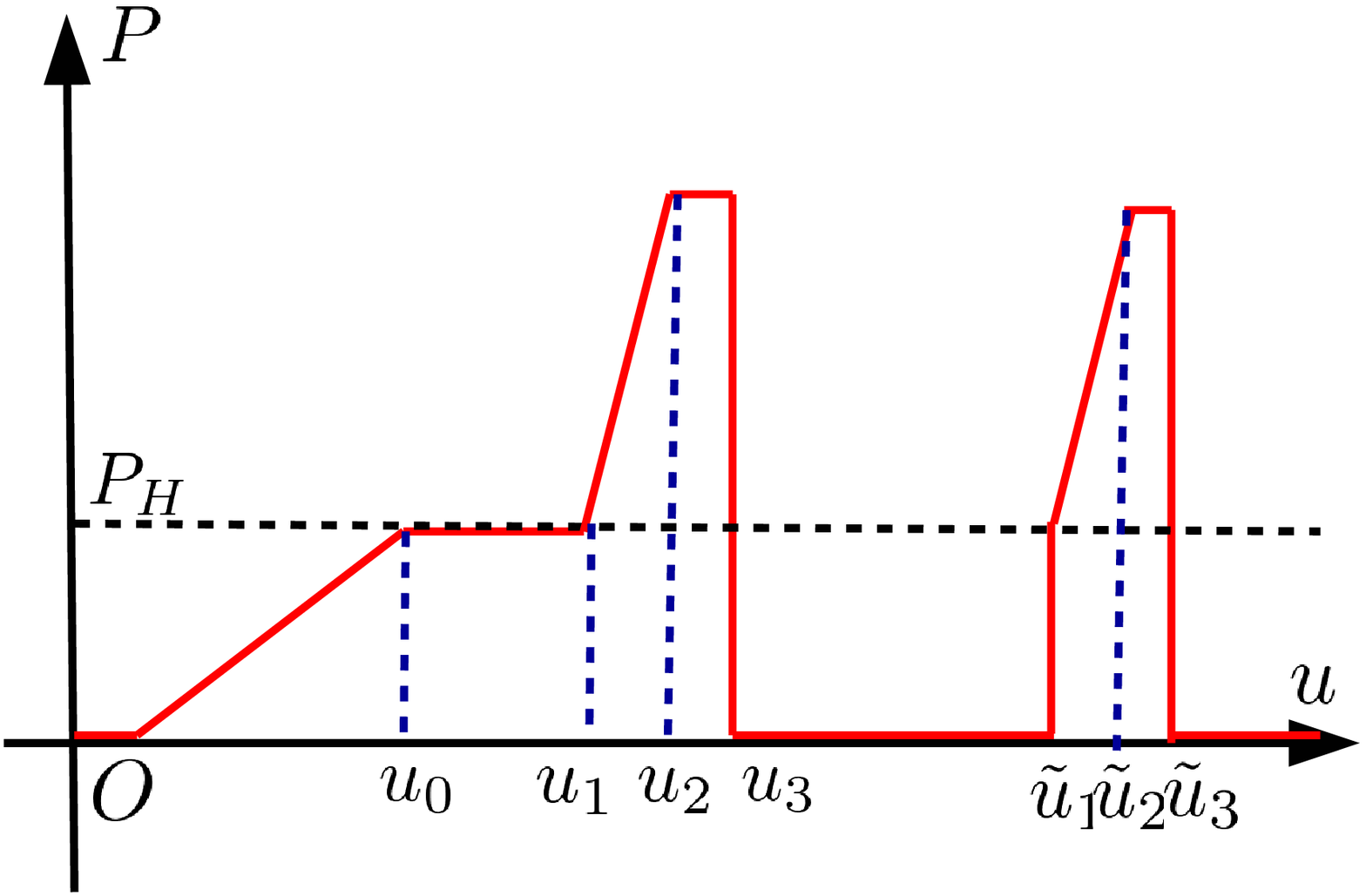}}
& 
\subfigure[
$\beta=1$]{\includegraphics[width=0.4\textwidth]{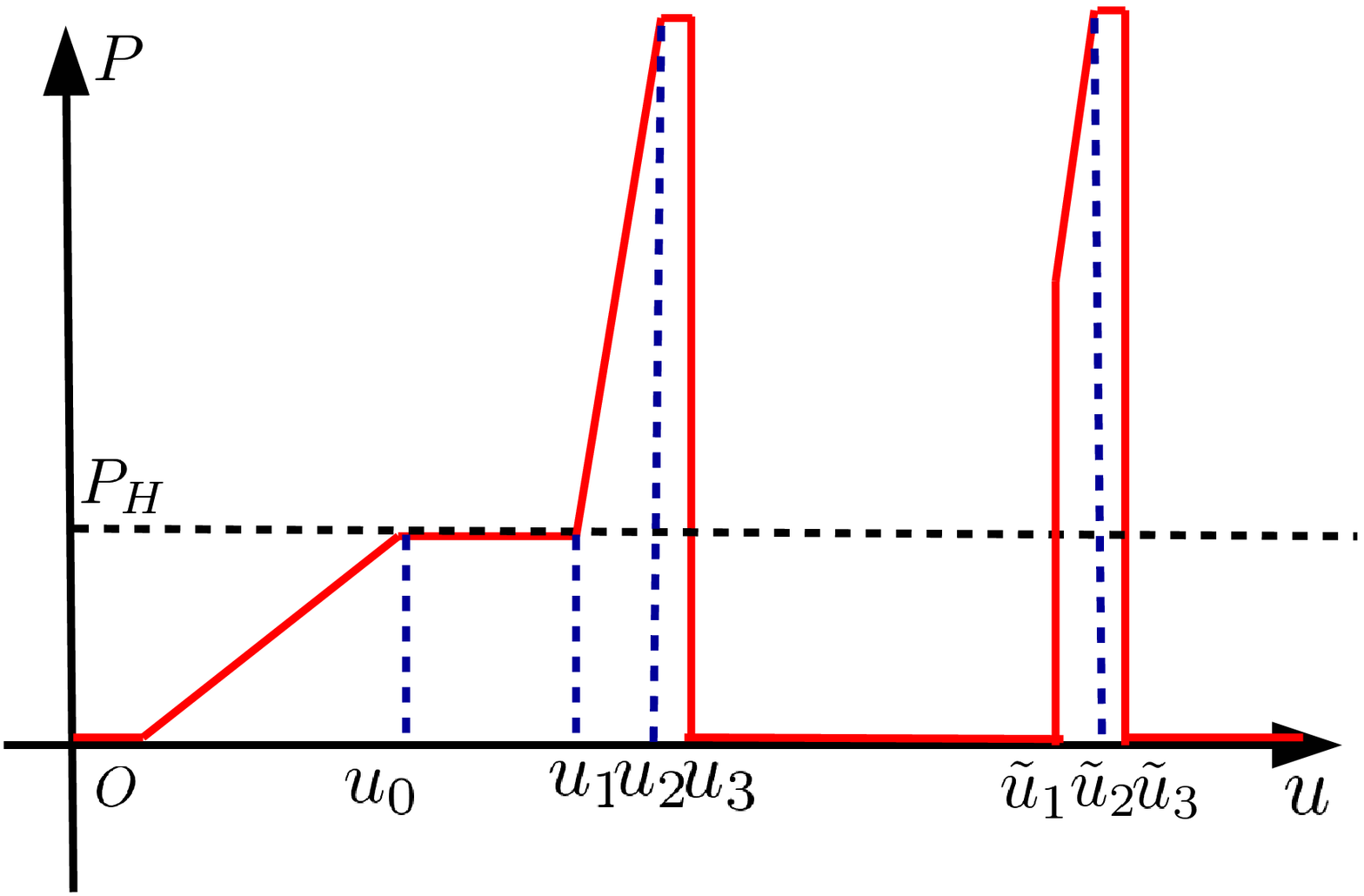}}
\end{tabular}
\caption{\label{fg:power_evolution_const} Same as
 Fig.~\ref{fg:power_evolution} but in model B, in which 
 the deceleration is constant in the braking phase.}
\end{center}
\end{figure}

For the post-Hawking burst, $\kappa$ decreases from $O( (2M)^{-1})$ 
to $ -1/(4M) \epsilon^{-2\beta}$
for $u_{1}<u<u_{2}$ and keeps constant with $\kappa\simeq
-1/(4M)\epsilon^{-2\beta}$ for $u_{2}<u<u_{3}$. 
For the late-time burst, $\kappa$ decreases from
$-1/(4M)\epsilon^{-(2\beta-1)}$ to $-1/(4M)\epsilon^{-2\beta}$ for 
$\tilde{u}_{1}<u<\tilde{u}_{2}$ and keeps constant with 
$\kappa\simeq - 1/(4M)\epsilon^{-2\beta}$ for $\tilde{u}_{2}<u<\tilde{u}_{3}$.
The power and energy radiated during the post-Hawking burst for
$u_{2}<u<u_{3}$ and the late-time burst for
$\tilde{u}_{2}<u<\tilde{u}_{3}$ are approximately the same in order of magnitude as
\begin{equation}
 P\simeq \epsilon^{-4\beta} P_{H}~~\mbox{and}~~E\simeq \frac{1}{48\pi}\frac{1}{8M}\epsilon^{-2\beta}
\end{equation}
for the duration $u_{3}-u_{2}\simeq \tilde{u}_{3}-\tilde{u}_{2} \simeq
2M\epsilon^{2\beta}$ as discussed in Appendix~\ref{sec:duration}.
Both of the bursts dominate the transient Hawking radiation in both
power and energy. The evolution of radiation is summarized for
$\beta=1/2$ and $\beta=1$ in Fig.~\ref{fg:power_evolution_const}.

\subsubsection{Instant deceleration limit}
It is interesting to see the limit $\beta\to \infty$ or $R_{b}-R_{f}\to
0$ while $1-(2M/R_{f})=\epsilon^{2}$ is fixed 
in both models A and B. In this limit, the power becomes
stronger and stronger, the time width becomes shorter and shorter, and 
the energy radiated becomes more and more in both the post-Hawking
and late-time bursts, 
while the duration of the dormant stage in between is unchanged. 
Thus, we can reproduce the last two delta-functional bursts
in the null-shell model in Sec.~\ref{sec:null_shell_model}.

\section{Discussion}
\label{sec:discussion}

It is important to compare our result with previous results in similar setups.
In Refs.~\cite{Paranjape:2009ib,Banerjee:2009nn}, a timelike-shell model was also used,
the end state of which is a static shell with radius slightly larger than $2M$. However, instead of prescribing the shell dynamics,
the function $G(u)$ was assumed directly to satisfy the expected qualitative asymptotic properties and change in a timescale of the order of
$M$. Figures 7 and 9 in Refs.~\cite{Paranjape:2009ib,Banerjee:2009nn}
indicate that the width of the late-time burst is several tens of $M$ and the power is bounded by that of the Hawking radiation $P_{H}$.
It was also observed that the width of the burst increases for smaller $\epsilon$.
As seen in Sec.~\ref{subsec:specific_models}, these features correspond
to our model A with $\beta=1/2$.
On the other hand, we can argue that the physically natural scenario
corresponds to model A with $\beta=1$ from the
argument that the shell begins to brake when $R=R_{b}=R_{f}+ 2M \epsilon^{2}$
and settles down to $R=R_{f}=2M/(1-\epsilon^{2})$, if there is a unique characteristic scale which
controls both the braking and the freeze-in of the shell and that
the force onto the shell is vanishingly small near $R=R_{f}$,
if $R= R_{f}$ is the radius of equilibrium. 

It is interesting to estimate the quantities which appear here using astrophysical values. 
The transient Hawking radiation lasts for $u_{1}-u_{0}\sim 40
 (M/M_{\odot}) [44+\ln (10^{-19}/\epsilon)]\,~\mu\mbox{s}$. The
radiation itself carries a power, temperature, and energy
\begin{eqnarray}
P&\simeq&  P_{H} \sim
 10^{-21}\left(\frac{M}{M_{\odot}}\right)^{-2}\,~\mbox{erg}/\mbox{s}\,,
 \quad T_{H}\sim  6\times 10^{-8}
 \left(\frac{M}{M_{\odot}}\right)^{-1}\,~\mbox{K}\,,\\
E&\simeq& 4\times
 10^{-26} \left(\frac{M}{M_{\odot}}\right)^{-1}\left[44+\ln \left(\frac{10^{-19}}{\epsilon}\right)\right]~\mbox{erg}.
\end{eqnarray}
The subsequent ``dormant'' stage lasts for $\tilde{u}_{1}-u_{3}\sim 6 \times 10^{6} (M/M_{\odot})\left(\epsilon/10^{-19}\right)^{-1}$ yr, and is followed by a late-time burst whose details depend on the model. 

For model A with $\beta=1/2$, the late-time burst lasts for $\tilde{u}_{2}-\tilde{u}_{1}\sim  2M(\ln \epsilon^{-1})\sim 10
(M/M_{\odot})[44+\ln (10^{-19}/\epsilon)]\,~\mu\mbox{s}$ and is
characterized by $P\sim  P_{H}$, $T=T_{H}$, and $E\sim E_{H}$.
For model A with $\beta=1$, 
the late-time burst lasts for $\tilde{u}_{2}-\tilde{u}_{1}\sim
10^{-24} (M/M_{\odot})\left(\epsilon/10^{-19}\right)[44+\ln
(10^{-19}/\epsilon)]\,~\mbox{s}$ and carries power, equivalent
temperature $T_{\rm eq}$, effective temperature $T_{\rm eff}$, and energy,
\begin{eqnarray}
P&\sim&
 10^{17}\left(\frac{\epsilon}{10^{-19}}\right)^{-2}\left(\frac{M}{M_{\odot}}\right)^{-2}\,~\mbox{erg}/\mbox{s}\,, \\
 kT_{\rm eq}&\sim &
 -100\left(\frac{\epsilon}{10^{-19}}\right)^{-1}\left(\frac{M}{M_{\odot}}\right)^{-1}~\mbox{MeV}, \\
 kT_{\rm eff}&\sim& 0.1~\left(\frac{\epsilon}{10^{-19}}\right)^{-1/2}\left(\frac{M}{M_{\odot}}\right)^{-1}~\mbox{eV}, \\
E&\sim&
10^{-7}\left(\frac{\epsilon}{10^{-19}}\right)^{-1}\left(\frac{M}{M_{\odot}}\right)^{-1}\left[44+\ln\left(10^{-19}/\epsilon\right)\right]\mbox{erg},
\end{eqnarray}
where $kT_{\rm eq}:=\kappa/(2\pi)$ and $kT_{\rm eff}:=(P/(4\pi
R^{2}\sigma_{\rm SB}/2))^{1/4}$ with $\sigma_{\rm SB}=\pi^{2}k^{4}/60$ the Stefan-Boltzmann constant,
while the post-Hawking burst carries approximately the same power
and same equivalent and effective temperatures.
For model B with $\beta=1$, these observables for both the first
and the second bursts are given by
\begin{eqnarray}
P&\sim& 
 10^{55}\left(\frac{\epsilon}{10^{-19}}\right)^{-4}\left(\frac{M}{M_{\odot}}\right)^{-2}\,~\mbox{erg}/\mbox{s}\,, \\
 kT_{\rm eq}&\sim &
 -10^{18}\left(\frac{\epsilon}{10^{-19}}\right)^{-2}\left(\frac{M}{M_{\odot}}\right)^{-1}~\mbox{GeV}, \\
 kT_{\rm
 eff}&\sim& 10~\left(\frac{\epsilon}{10^{-19}}\right)^{-1}\left(\frac{M}{M_{\odot}}\right)^{-1}~\mbox{MeV}, \\
E&\sim& 
10^{12}\left(\frac{\epsilon}{10^{-19}}\right)^{-2}\left(\frac{M}{M_{\odot}}\right)^{-1}\mbox{erg},
\end{eqnarray}
with time widths $u_{3}-u_{2}\simeq \tilde{u}_{3}-\tilde{u}_{2}\simeq
10^{-43}~(\epsilon/10^{-19})^{2}$ s.

Here, we would like to discuss some remaining issues. The first concerns arguments for the ``expected'' values of our $\epsilon$ parameter.
Although $\epsilon\sim \sqrt{\ell_{Pl}/(2M)}\simeq 10^{-19} (M/M_{\odot})^{-1/2}$ is suggested by some semiclassical arguments~\cite{Mazur:2004fk},
other scenarios where $\epsilon$ can be much larger or smaller than this value are possible. For example, 
one might identify the proper length from the surface with the Planck length~\footnote{The authors are grateful to T.~Tanaka for pointing out this possibility.} (instead of the areal radius). In such a case, $\epsilon$ can be as small as $\epsilon\simeq l_{P}/(4M)\simeq 10^{-38}
(M/M_{\odot})^{-1}$ and our results become even more extreme. On the other hand, if we consider a neutron star, we may estimate
$\epsilon\sim 0.5$, for which the present formulation is only marginally valid. 

We have shown that the duration of the dormant stage is $\sim 4M/\epsilon$. Physically, the $4M$ factor is simply the proper time of the shell 
for a null ray to cross its diameter, when the shell is sufficiently close to $2M$. 
The factor $1/\epsilon$ comes from the redshift factor between the proper time of the almost static shell and the observer time.

The particle production process is characterized by different stages,
after what we termed the ``standard'' collapse phase.
This large number of particle production stages is due to 
the different classes of null-ray pairs that govern quantum particle creation. We can summarize the correspondence as
follows: braking at $\tau=\tau_{\rm out}$ and standard collapse at $\tau=\tau_{{\rm in}}$ contribute to the post-Hawking burst, the final static phase at $\tau=\tau_{{\rm out}}$ and standard collapse at $\tau=\tau_{{\rm in}}$ produce the dormant stage, whereas the final static phase at $\tau=\tau_{{\rm out}}$ and braking at $\tau=\tau_{{\rm in}}$ give the
late-time burst.

We have applied the geometrical optics approximation in the entire treatment. This is valid for $s$-waves and for sufficiently high frequencies. On the other hand, the reflection of waves 
by the shell and the geometry is completely neglected. This implies that if we relax this approximation, we will obtain
not only the post-Hawking and late-time bursts but also echoes in
particle creation due to the reflections of waves (cf. Refs.~\cite{Cardoso:2016rao,Cardoso:2017cqb,Cardoso:2017njb}). 
The details of this process require further calculations.

As we pointed out, we adopted the same set of assumptions
for calculating quantum particle creation from a 
collapsing body
as previous works in the 
literature~\cite{Hawking:1974rv,Hawking:1974sw,Ford:1978ip}.
It is clearly important to go beyond such restrictions.
If one goes beyond the geometrical optics approximation,
Hawking radiation appears as a stationary process at 
the final stage of the collapse to a black hole with 
various intermediate decaying stages~\cite{Akhmedov:2015xwa}.
Quantum loop corrections to the flux of non-Gaussian (self-interacting)
theories are not suppressed in comparison with the tree-level
contribution in the case of 
$\lambda\phi^{4}$ theory~\cite{Akhmedov:2015xwa}. 
This may also modify the properties 
of the Hawking radiation and perhaps those of the bursts 
discussed in the current paper.
Furthermore, the properties of radiation in the intermediate stage 
of the collapse may strongly depend on the choice of the initial 
quantum state.

Finally, we have prescribed the shell dynamics in this paper, 
but postpone a discussion about the matter content of the
shell which enables such an unusual time evolution. We expect that some energy conditions must be violated. 
The physical significance of such violations is not completely clear. However, we take this opportunity to once more stress
that one of the main goals of this work is to look for distinctive features of horizonless objects as a way to strengthen the black hole paradigm.

\acknowledgments
The authors are grateful to T.~Igata, S.~Kinoshita, H.~Maeda,
A.~Matsumura, F.~C.~Mena, K.~i.~Nakao, 
K.~Nakashi, J.~M.~M.~Senovilla, and T.~Tanaka for helpful comments and fruitful discussions.
This work was partially supported by JSPS KAKENHI Grant No. JP26400282
(T.H.). T. H. thanks CENTRA, IST, Lisbon, for hospitality received during this work.
The authors are grateful to Yukawa Institute for Theoretical Physics at Kyoto University for hospitality while this work was completed during the YITP-T-18-05 on ``Dynamics in Strong Gravity Universe.'' 
T. H. was supported by Rikkyo University International Academic Research Exchange.
The authors acknowledge financial support provided under the European Union's H2020 ERC Consolidator Grant ``Matter and strong-field gravity: New frontiers in Einstein's theory'' Grant Agreement No. MaGRaTh--646597. 
The authors acknowledge financial support provided by FCT/Portugal
through Grant No. PTDC/MAT-APL/30043/2017.
This work has received funding from the European Union's Horizon 2020 research and innovation program under the 
Marie Sk\l odowska-Curie Grant Agreement No. 690904, and through NSF-XSEDE Award No. PHY-090003.
This article is based upon work from COST Action CA16104 ``GWverse,'' supported by COST (European Cooperation in Science and Technology).


\appendix
\section{Expressions for a timelike-shell model \label{sec:ABCD_general}}
The junction condition for the first fundamental form gives
\begin{equation}
\dot{T}^{2}=1+\dot{R}^{2}\,,\quad 
\dot{t}^{2}=\frac{1}{1-\frac{2M}{R}}\left(1+\frac{\dot{R}^{2}}{1-\frac{2M}{R}}\right)\,.
\end{equation}
The relation between the null coordinates and the proper time of the
shell is given by
\begin{equation}
 \dot{U}=\sqrt{1+\dot{R}^{2}}-\dot {R},\quad 
 \dot{V}=\sqrt{1+\dot{R}^{2}}+\dot {R}\,,\label{eq:U_dot_V_dot}
\end{equation}
and 
\begin{equation}
 \dot{u}=\frac{\sqrt{1-\frac{2M}{R}+\dot{R}^{2}}-\dot {R}}{1-\frac{2M}{R}},\quad 
 \dot{v}=\frac{\sqrt{1-\frac{2M}{R}+\dot{R}^{2}}+\dot {R}}{1-\frac{2M}{R}}\,.\label{eq:u_dot_v_dot}
\end{equation}

From Eqs.~(\ref{eq:U_dot_V_dot}) and (\ref{eq:u_dot_v_dot}), we 
can write down the explicit expression 
for $A$ and $B$ in terms of $R$ as follows:
\begin{eqnarray}
A=
\frac{\left(1-\frac{2M}{R}\right)\left(\sqrt{1+\dot{R}^{2}}-\dot{R}\right)}{\sqrt{1-\frac{2M}{R}+\dot{R}^{2}}-\dot{R}}\,,
\quad 
B=
\frac{\left(1-\frac{2M}{R}\right)\left(\sqrt{1+\dot{R}^{2}}+\dot{R}\right)}{\sqrt{1-\frac{2M}{R}+\dot{R}^{2}}+\dot{R}}\,.
\label{eq:A_B_explicit}
\end{eqnarray}
From Eqs.~(\ref{eq:u_dot_v_dot}) and (\ref{eq:A_B_explicit}), 
we can write down the expression for $C$ and $D$ in terms of $R$ as follows:
\begin{eqnarray}
 C&=&-\frac{\ddot{R}\left(1-\frac{2M}{R}\right)}{\sqrt{1-\frac{2M}{R}+\dot{R}^{2}}-\dot{R}}
\left[\frac{1}{\sqrt{1-\frac{2M}{R}+\dot{R}^{2}}}-\frac{1}{\sqrt{1+\dot{R}^{2}}}\right]-\frac{M\dot{R}}{R^{2}\sqrt{1-\frac{2M}{R}+\dot{R}^{2}}}\,, \\
D&=&\frac{\ddot{R}\left(1-\frac{2M}{R}\right)}{\sqrt{1-\frac{2M}{R}+\dot{R}^{2}}+\dot{R}}
\left[\frac{1}{\sqrt{1-\frac{2M}{R}+\dot{R}^{2}}}-\frac{1}{\sqrt{1+\dot{R}^{2}}}\right]-\frac{M\dot{R}}{R^{2}\sqrt{1-\frac{2M}{R}+\dot{R}^{2}}}\,.
\label{eq:C_D_explicit}
\end{eqnarray}
%

\section{Expressions for a timelike-shell model in different regimes
 \label{sec:ABCD}}
To estimate the functions $A$, $B$, $C$, and $D$, 
we are interested in the following phases: 0. ($R\gg 2M$ and
$|\dot{R}|\ll 1$), 1., 2. ($1-\frac{2M}{R}\ll 1$ and $1-\frac{2M}{R} \ll \dot{R}^{2}$),
 3. ($1-\frac{2M}{R}\ll 1$ and $1-\frac{2M}{R} \ll \dot{R}^{2}$), and 4. ($R=\mbox{const}$). Let us consider these cases separately.
We assume $\dot{R}<0$ in the following.

\begin{flushleft}
{\bf\quad 0. $R\gg 2M$ and $|\dot{R}|\ll 1$} 
\end{flushleft}
We find
\begin{eqnarray}
A\simeq   1\,, ~~
B\simeq  1\,, ~~
C\simeq  -\frac{M}{R}\left(\ddot{R}-\frac{|\dot{R}|}{R}\right)\,,~~
D\simeq  \frac{M}{R}\left(\ddot{R}+\frac{|\dot{R}|}{R}\right)\,.
\label{eq:1_ABCD}
\end{eqnarray}
From Eqs.~(\ref{eq:U_dot_V_dot}) and
(\ref{eq:u_dot_v_dot}), we obtain
\begin{equation}
 U\simeq \tau+\mbox{const},\quad V\simeq \tau+\mbox{const},\quad 
 u\simeq \tau+\mbox{const},\quad v\simeq \tau+\mbox{const}.
\end{equation}

\begin{flushleft}
{\bf\quad 1. and 2. $1-\frac{2M}{R}\ll 1$ and $1-\frac{2M}{R} \ll \dot{R}^{2}$}
\end{flushleft}
In this regime, we have
\begin{eqnarray}
A&\simeq &
 \left(1-\frac{2M}{R}\right)\frac{\sqrt{1+\dot{R}^{2}}+|\dot{R}|}{2|\dot{R}|}\,,
 \label{eq:2_A} \\
B&\simeq & 2|\dot{R}|(\sqrt{1+\dot{R}^{2}}-|\dot{R}|)\,, \label{eq:2_B}\\
C&\simeq & -\frac{\ddot{R}}{2\dot{R}^{2}}\left(1-\frac{2M}{R}\right)\left[1-\frac{|\dot{R}|}{\sqrt{1+\dot{R}^{2}}}\right]+\frac{1}{4M}\,, \label{eq:2_C}\\
D&\simeq &
2\ddot{R}\left[1-\frac{|\dot{R}|}{\sqrt{1+\dot{R}^{2}}}\right]+\frac{1}{4M}\, 
\label{eq:2_D}.
\end{eqnarray}
In this case, from Eqs.~(\ref{eq:u_dot_v_dot}), $u$ and $v$ are given by 
\begin{equation}
 u\simeq -4M\ln \left[\frac{R}{2M}-1\right]+\mbox{const},\quad 
 v\simeq -\frac{1}{2}\int \frac{dR}{\dot{R}^{2}}.
\label{eq:u_v_R_late-collapse} 
\end{equation}

If we further assume $\dot{R}=O(1)$ and $\ddot{R}=O((2M)^{-1})$
corresponding to phase 1, we obtain 
\begin{equation}
 A=O\left(\left(1-\frac{2M}{R}\right)\right), ~~
 B=O(1),~~C\simeq \frac{1}{4M},~~
 D=O((2M)^{-1}).
\label{eq:late-collapse_phase}
\end{equation}
In this case, Eqs.~(\ref{eq:U_dot_V_dot}) imply
\begin{equation}
 U\sim \tau+\mbox{const},\quad 
 V\sim \tau+\mbox{const}.
\end{equation}

\begin{flushleft}
{\bf\quad 3. $1-\frac{2M}{R}\ll 1$ and $1-\frac{2M}{R} \gg \dot{R}^{2}$}
\end{flushleft}

In this regime, we have
\begin{equation}
 A\simeq \sqrt{1-\frac{2M}{R}},\quad B\simeq \sqrt{1-\frac{2M}{R}},\quad  
C\simeq -\ddot{R}+\frac{|\dot{R}|}{4M\sqrt{1-\frac{2M}{R}}}, 
\quad
D\simeq \ddot{R}+\frac{|\dot{R}|}{4M\sqrt{1-\frac{2M}{R}}}.
    \end{equation}
In this case, from Eqs.~(\ref{eq:U_dot_V_dot}) and 
(\ref{eq:u_dot_v_dot}), we obtain
\begin{equation}
 U\simeq \tau+\mbox{const},\quad 
 V\simeq \tau+\mbox{const},\quad 
 u\simeq \int \frac{d\tau}{\sqrt{1-\frac{2M}{R}}},\quad 
 v\simeq \int \frac{d\tau}{\sqrt{1-\frac{2M}{R}}}.
\label{eq:u_v_tau_late-brakling_static}
\end{equation}

\begin{flushleft}
{\bf\quad 4. $R=\mbox{const}$}
\end{flushleft}
In this regime, we have
\begin{eqnarray}
A=B=\sqrt{1-\frac{2M}{R}},\quad C=D=0,
\end{eqnarray}
and 
\begin{equation}
 U=\tau+\mbox{const},~~V=\tau+\mbox{const},~~
 u=\frac{\tau}{\sqrt{1-\frac{2M}{R}}}+\mbox{const},~~
 v=\frac{\tau}{\sqrt{1-\frac{2M}{R}}}+\mbox{const}.
\end{equation}

\section{Time intervals\label{sec:duration}}

Since $R=4M$ at $u=u_{0}$ and 
$R=R_{b}=R_{f}+2M\epsilon^{2\beta}$ at $u=u_{1}$,
Eq.~(\ref{eq:u_v_R_late-collapse}) implies 
\begin{eqnarray}
u_{1}-u_{0} \simeq 
\left\{\begin{array}{cc}
		     4M\beta\ln \epsilon^{-2} & (0< \beta<1)\\
		     4M\ln \epsilon^{-2} & (\beta\ge 1) 
			   \end{array}\right. .
\end{eqnarray}
Equations~(\ref{eq:U_dot_V_dot}) and (\ref{eq:u_dot_v_dot}) imply that the intervals in terms of $u$ are given as follows:
\begin{eqnarray}
\tilde{u}_{2}-u_{2} &\simeq & \epsilon^{-1}
      (\tilde{\tau}_{2}-\tau_{2})\simeq \frac{4M}{\epsilon}, \\
\tilde{u}_{3}-u_{3} &\simeq & \epsilon^{-1}
      (\tilde{\tau}_{3}-\tau_{3})\simeq \frac{4M}{\epsilon}, 
\end{eqnarray}
where we have used $\tilde{\tau}_{2}-\tau_{2}
\simeq \tilde{\tau}_{3}-\tau_{3} \simeq 4M$. The above relations do not
depend on the details of the model.

\subsection{Model A: Exponentially slowed-down model}
First, we estimate $\tau_{2}$. Assuming $R-R_{f}\propto
e^{-\sigma\tau}$ for $\tau_{1}<\tau<\tau_{3}'$ with $\sigma$ given by
Eq.~(\ref{eq:sigma}), 
we find
\begin{equation}
 1-\frac{2M}{R}\simeq \left(1-\frac{2M}{R_{f}}\right)+\left(\frac{2M}{R_{f}}-\frac{2M}{R}\right)\simeq \epsilon^{2}+\epsilon^{2\beta}e^{-\sigma(\tau-\tau_{1})}
\end{equation}
for $R_{b}-R_{f}=2M\epsilon^{2\beta}$ with $\beta \ge 1/2$.
Noting $|\dot{R}|=|\dot{R}_{b}|e^{-\sigma(\tau-\tau_{1})}$
and $|\dot{R}_{b}|=O(1)$, 
Eq.~(\ref{eq:hierarchy_marginal}) at $\tau=\tau_{2}$ implies  
\begin{equation}
 \tau_{2}-\tau_{1}\simeq \sigma^{-1}(\ln \epsilon^{-1})~~\mbox{and}
 ~~ 1-\frac{2M}{R_{2}}\simeq \epsilon^{2}.
\end{equation}

Then, we can derive 
\begin{eqnarray}
u_{2}-u_{1} &\simeq&  
\left\{\begin{array}{cc}
		     4M(1-\beta)\ln \epsilon^{-2} & (1/2\le  \beta<1)\\
		     4M & (\beta=1) \\
		      4M\epsilon^{2(\beta-1)} & (\beta>1)
			   \end{array}\right. ,  \\
u_{2}-u_{0}&\simeq& 4M\ln \epsilon^{-2}.
\end{eqnarray}
Equations~(\ref{eq:U_dot_V_dot}) and (\ref{eq:u_dot_v_dot}) imply that the intervals in terms of $u$ are given as follows:
\begin{eqnarray}
\tilde{u}_{2}-\tilde{u}_{1}&\simeq
 &\epsilon^{-1}(\tilde{\tau}_{2}-\tilde{\tau}_{1}) 
\sim \epsilon^{-1}(\tau_{2}-\tau_{1})\sim \epsilon^{-1}(\ln
\epsilon^{-1})\sigma^{-1} \sim
		     2M\epsilon^{2\beta-1}(\ln \epsilon^{-1}),
\label{eq:tilde_u_2-tilde_u_1}
\end{eqnarray}
where we have used $\sigma\simeq 1/(R_{b}-R_{f})$. Additionally 
assuming $\tau_{3}-\tau_{2}\simeq \sigma^{-1}$, we can find
\begin{eqnarray}
 u_{3}-u_{2}\simeq \tilde{u}_{3}
-\tilde{u}_{2}\simeq \epsilon^{-1}\sigma^{-1}\simeq 2M\epsilon^{2\beta-1}.
\end{eqnarray}
\subsection{Model B: Constant-deceleration model}

In this model, we find 
\begin{equation}
 1-\frac{2M}{R}\simeq \epsilon^{2}+\frac{a}{4M}(\tau_{3}-\tau)^{2}
\end{equation}
for $\tau_{1}<\tau<\tau_{3}$ with $a$ given by Eq.~(\ref{eq:a}).
Noting $|\dot{R}|=a(\tau_{3}-\tau)$ and $|\dot{R}_{b}|=O(1)$,  
$\tau_{2}$ is estimated as
\begin{equation}
 \tau_{3}-\tau_{2}=\frac{\epsilon}{\sqrt{a\left(a-\frac{1}{4M}\right)}},
\end{equation}
while $\tau_{3}$ is estimated as $\tau_{3}-\tau_{1}=|\dot{R}_{b}|
a^{-1}$. 
Then, we can derive 
\begin{eqnarray}
u_{2}-u_{1} &\simeq&  
\left\{\begin{array}{cc}
		     4M(1-\beta)\ln \epsilon^{-2} & (0<  \beta<1)\\
		     4M & (\beta=1) \\
		      4M\epsilon^{2(\beta-1)} & (\beta>1)
			   \end{array}\right. ,  \\
u_{2}-u_{0}&\simeq& 4M\ln \epsilon^{-2}.
\end{eqnarray}
The expressions for $u_{3}-u_{2}$,
$\tilde{u}_{3}-\tilde{u_{1}}$ and $\tilde{u}_{3}-\tilde{u}_{2}$ are
given by 
\begin{eqnarray}
 u_{3}-u_{2}&\simeq& \epsilon^{-1}(\tau_{3}-\tau_{2})\simeq a^{-1}\simeq
  4M\epsilon^{2\beta}, \\
 \tilde{u}_{3}-\tilde{u}_{1}&\simeq &
  \epsilon^{-1}(\tau_{3}-\tau_{1})\simeq
  |\dot{R}_{b}|\epsilon^{-1}a^{-1}\simeq 4M\epsilon^{2\beta-1}, \\
 \tilde{u}_{3}-\tilde{u}_{2}&\simeq &
  \epsilon^{-1}(\tau_{3}-\tau_{2})\simeq a^{-1}\simeq 4M\epsilon^{2\beta},
\end{eqnarray}
where we have assumed $a\gg 1/(4M)$ and $a\simeq 1/[2(R_{b}-R_{f})]=1/(4M\epsilon^{2\beta})$.

\section{Detailed analysis of the temporal change of radiation
 \label{sec:radiative_histories}}
We divide the observer's time to eight intervals: $u<u_{0}$,
$u_{0}<u<u_{1}$, $u_{1}<u<u_{2}$, $u_{2}<u<u_{3}$, $u_{3}<u<\tilde{u}_{1}$,
$\tilde{u}_{1}<u<\tilde{u}_{2}$, $\tilde{u}_{2}<\tilde{u}_{3}$, and
$\tilde{u}_{3}<u$. 
Since the first two are identical to those in the standard-collapse phase,
discussed in Sec.~\ref{sec:standard_collapse}, we concentrate on the
last six. For each interval, the classes of null-ray pairs are
fixed and we can obtain the expressions for the functions $G'(u)$ and
$\kappa(u)$ by combining the expressions for $A$, $B$, $C$ and $D$
given in Appendix~\ref{sec:ABCD}
through the formula (\ref{eq:general_formula_for_G'_kappa}).
\begin{itemize}

\item[(i)] $u_{1}<u<u_{2}$\\
We discuss this stage in Sec.~\ref{subsec:post-Hawking_burst}.
There are null-ray pairs of classes $(2,1)$ and $(2,0)$. There is no
pair of class $(2,2)$ because the duration of phase 2, $\tau_{2}-\tau_{1}$, is much shorter than the time for return travel, which is approximately $4M$. 

\item[(ii)] $u_{2}<u<u_{3}$\\
We also discuss this stage in Sec.~\ref{subsec:post-Hawking_burst}.
We have null-ray pairs of classes $(3,0)$ and $(3,1)$. 
There is no pair of class $(3,2)$ or $(3,3)$ 
because we assume that $\tau_{3}-\tau_{2}$ is much shorter than $4M$.

\item[(iii)] $u_{3}<u<\tilde{u}_{1}$\\
All null-ray pairs are of class $(4,0)$ or $(4,1)$. We can find 
      $\kappa=O(\epsilon (2M)^{-1})$, which 
      is the contribution from $\tau=\tau_{{\rm in}}$ in phases 0 and
      1, irrespectively of the model details. This corresponds to the
      long dormant stage.  
\item[(iv)] $\tilde{u}_{1}<u<\tilde{u}_{2}$\\ 
We discuss this stage in Sec.\ref{subsec:late-time_burst}.
The null-ray pairs are of class $(4,2)$.

\item[(v)] $\tilde{u}_{2}<u<\tilde{u}_{3}$\\
We also discuss this stage in Sec.\ref{subsec:late-time_burst}.
We have null-ray pairs of class $(4,3)$. 

\item[(vi)] $\tilde{u}_{3}<u$\\
We have null-ray pairs of class $(4,4)$. For this class, we have just 
$G'(u)=  1\,$ and $\kappa(u)= 0$.
The radiation completely vanishes at $u=\tilde{u}_{3}$ and thereafter 
no radiation is emitted forever.
\end{itemize}


\end{document}